\documentclass[authoryear, review, times]{elsarticle}
\usepackage{natbib}
\usepackage{graphicx}
\usepackage{amssymb}
\usepackage{url}

\journal{Planetary and Space Sciences}

\begin{document}

\begin{frontmatter}

\title{Escape of the martian protoatmosphere and\\ initial water inventory}

\author[icm]{N. V. Erkaev\fnref{sib}}
\author[iwf]{H. Lammer\corref{cor1}}
\ead{helmut.lammer@oeaw.ac.at}
\author[cis]{L. Elkins-Tanton}
\author[ifa]{A. St\"{o}kl}
\author[iwf]{P. Odert\fnref{igam}}
\author[lat]{E. Marcq}
\author[ifa]{E. A. Dorfi}
\author[iwf]{K. G. Kislyakova}
\author[pgi]{Yu.~N.~Kulikov}
\author[igam]{M.~Leitzinger}
\author[ifa]{M. G\"{u}del}

\address[icm]{Institute for Computational Modelling, 660041 Krasnoyarsk 36, Russian Academy of Sciences, Russian Federation}
\address[sib]{Siberian Federal University, 660041 Krasnoyarsk, Russian Federation}
\address[iwf]{Space Research Institute, Austrian Academy of Sciences, Schmiedlstrasse 6, A-8042 Graz, Austria}
\address[cis]{Department of Terrestrial Magnetism, Carnegie Institution for Science, Washington DC 20015, USA}
\address[ifa]{Institute for Astronomy, University of Vienna, T\"{u}rkenschanzstra{\ss}e 17  1180 Vienna, Austria}
\address[lat]{LATMOS, Universit\'{e} de Versailles Saint-Quentin-en-Yvelines, Guyancourt, France}
\address[pgi]{Polar Geophysical Institute, Russian Academy of Sciences,\\ Khalturina 15, 183010 Murmansk, Russian Federation}
\address[igam]{Institute of Physics, IGAM, University of Graz, Universit\"atsplatz 5, A-8010  Graz, Austria}

\cortext[cor1]{Principal corresponding author}

\begin{abstract}
Latest research in planet formation indicate that Mars formed within a few million years (Myr) and remained a planetary embryo that never grew to a more massive planet.
It can also be expected from dynamical models, that most of Mars' building blocks consisted of material that formed in orbital locations just beyond the ice line which
could have contained $\sim$0.1--0.2 wt. \% of H$_2$O. By using these constraints, we estimate the nebula-captured and catastrophically outgassed volatile contents
during the solidification of Mars' magma ocean and apply a hydrodynamic upper atmosphere model for the study of the soft X-ray and extreme ultraviolet (XUV)
driven thermal escape of the martian protoatmosphere during the early active epoch of the young Sun.
The amount of gas that has been captured from the protoplanetary disk into the
planetary atmosphere is calculated by solving the hydrostatic structure equations in the protoplanetary nebula. Depending on nebular properties such as the
dust grain depletion factor, planetesimal accretion rates and luminosities, hydrogen envelopes with masses $\geq 3 \times 10^{19}$ g
to $\leq 6.5 \times 10^{22}$ g could have been captured from the nebula around early Mars. Depending of the before mentioned parameters,
due to the planets low gravity and a solar XUV flux that was $\sim$100 times stronger compared to the present value,
our results indicate that early Mars would have lost its nebular captured hydrogen envelope after the nebula gas evaporated,
during a fast period of $\sim$0.1 - 7.5 Myr. After the solidification of early Mars' magma ocean, catastrophically outgassed volatiles
with the amount of $\sim$50--250 bar H$_2$O and $\sim$10--55 bar CO$_2$ could have been lost during $\sim$0.4--12 Myr,
if the impact related energy flux of large planetesimals and small embryos to the planet's surface lasted long enough, that the steam atmosphere could have been
prevented from condensing. If this was not the case, then our results suggest that,
the timescales for H$_2$O condensation and ocean formation may have been shorter compared to the atmosphere evaporation timescale,
so that one can speculate that sporadically periods, where some amount of liquid water may have been present on the planet's surface.
However, depending on the amount of the outgassed volatiles, because of impacts and the high XUV-driven atmospheric escape rates,
such sporadically wet surface conditions may have not lasted longer than $\sim$0.4--12 Myr. After the loss of the captured hydrogen envelope
and outgassed volatiles
during the first 100 Myr period of the young Sun, a warmer and probably wetter
period may have evolved by a combination of volcanic outgassing and impact delivered volatiles $\sim 4.0\pm 0.2$ Gyr ago,
when the solar XUV flux decreased to values that have been $<$ 10 times that of today's Sun.
\end{abstract}


\end{frontmatter}

\section{Introduction}
The formation of Mars' nebula-captured, catastrophically degassed and impact delivered protoatmosphere is directly
connected to the planet's formation time scale, the nebula dissipation time, its orbital location and
the planet's small mass compared to Earth and Venus. Chassefi\`{e}re (1996a; 1996b) investigated for the first time the
hydrodynamic loss of oxygen from primitive atmospheres of Venus and Mars in detail. However, the pioneering
studies of Chassefi\`{e}re (1996a; 1996b) are based on meanwhile outdated terrestrial planet formation models in which the time
of the final accretion for terrestrial planets occurred $\geq$100 Myr after the formation of the Sun (Wetherill, 1986).
Furthermore, in these pioneering studies by Chassefi\`{e}re (1996a; 1996b) the cooling phase of the magma ocean
was expected to occur after $\sim$100 Myr, while more recent studies indicate that the solidification of magma oceans
even with depths of up to $\sim$2000 km is a fast process and mantle solidification of $\sim$98\% can be completed in
$\leq$5 Myr (e.g. Elkins-Tanton, 2008; Elkins-Tanton, 2011; Marcq, 2012; Lebrun et al., 2013; Hamano et al., 2013).
Moreover, it is also important to note that the assumption of several previous studies, that terrestrial planets, including early
Mars finished their accretion late, resulted also in ages where the soft X-ray and extreme ultraviolet (XUV) flux of the young
Sun was much lower compared to the high XUV flux values, which are now known from multi-wavelength observations of so-called
young solar proxies (e.g., G\"{u}del et al., 1997; Ribas et al., 2005; G\"{u}del, 2007; Claire et al., 2012).
Because of the lack of accurate data, Chassefi\`{e}re (1996a; 1996b) applied as its highest value an XUV enhancement
factor which was $\sim$25 times higher than that of the present Sun.

In a recent review article on Mars' origin Brasser (2013) argued that
Mars' small mass requires that the terrestrial planets have formed from a narrow annulus
of material, rather than a disc extending to Jupiter. The truncation of the outer part of the disc
was most likely related to migration of the gas giants, which kept the mass of Mars small. For the
formation of the martian protoatmosphere this evidence from planet formation and latest dynamical models
has important implications, because it would mean that Mars formed within a few million years and can be
considered as a planetary embryo that never grew to a ``real'' more massive planet. Moreover, from the latest
martian formation modeling scenarios most likely related to migration of the
giants (Walsh et al., 2011) it is expected that most of the planet's building blocks consists
of material that formed in a region just behind the ice line, so that the materials were more
water-rich than the materials that were involved in the accretion of Venus and Earth.

Brasser (2013) suggest that the building blocks of early Mars could have consisted of $\sim$0.1--0.2 wt.\% of H$_2$O.
The results presented in Brasser (2013) which are based on studies by Walsh et
al. (2011) agree in the amount of Mars' initial water inventory with Lunine et al. (2003) who applied also a
dynamical model which yielded longer formation time scales.
However, it should also be pointed out that model studies
which consider different impact regimes than the before mentioned studies can also result in an early Mars which
originated drier (Horner, 2009). Although, it is obvious that our current knowledge of terrestrial planet formation
and its related hydration is presently insufficient there is geomorphological evidence for water on early Mars,
where $\sim$90 \% was most likely outgassed and/or delivered during the first Gyr
(e.g., Chassefi\`{e}re, 1996b; 2013; Baker, 2001; Lammer et al., 2013a).

The main aim of the present study is to investigate in detail how long the before mentioned nebular captured and
catastrophically outgassed protoatmospheres have been stable after Mars' origin, to understand how long the early planet's
protoatmosphere survived against thermal atmospheric escape. In Sect. 2 the formation of a nebula captured hydrogen envelope on early Mars
and the expected catastrophically outgassed steam-type protoatmosphere based on materials which contain
$\sim$0.1--0.2 wt. \% H$_2$O (Brasser, 2013) is described. In Sect. 3 we discuss the early XUV radiation environment of the young
Sun and the life time of the nebula gas which determines the age when the planet's protoatmosphere was exposed freely to the high
solar XUV radiation field. In Sect. 4 we study the upper atmosphere structure and the
escape of the martian protoatmosphere by applying a time-dependent numerical algorithm, which is able to solve the
system of 1-D fluid equations for mass, momentum, and energy conservation. Finally we describe the solar and atmospheric
input parameters of the applied model and discuss the results.
\section{Nebula-based and catastrophically outgassed protoatmospheres}
For studying the potential habitability and atmosphere evolution of Mars, it is important to
understand which sources and sinks contributed to the formation of the planet's
initial atmosphere and water inventory. Furthermore, a detailed investigation on the
escape-related evolution of the early martian protoatmosphere is important for understanding
how long Mars may have had surface conditions that standing bodies of liquid water could have existed on the
planet's surface. Generally four main processes are responsible for the formation of planetary atmospheres
\begin{figure}[h]
\begin{center}
\includegraphics[width=0.95\columnwidth]{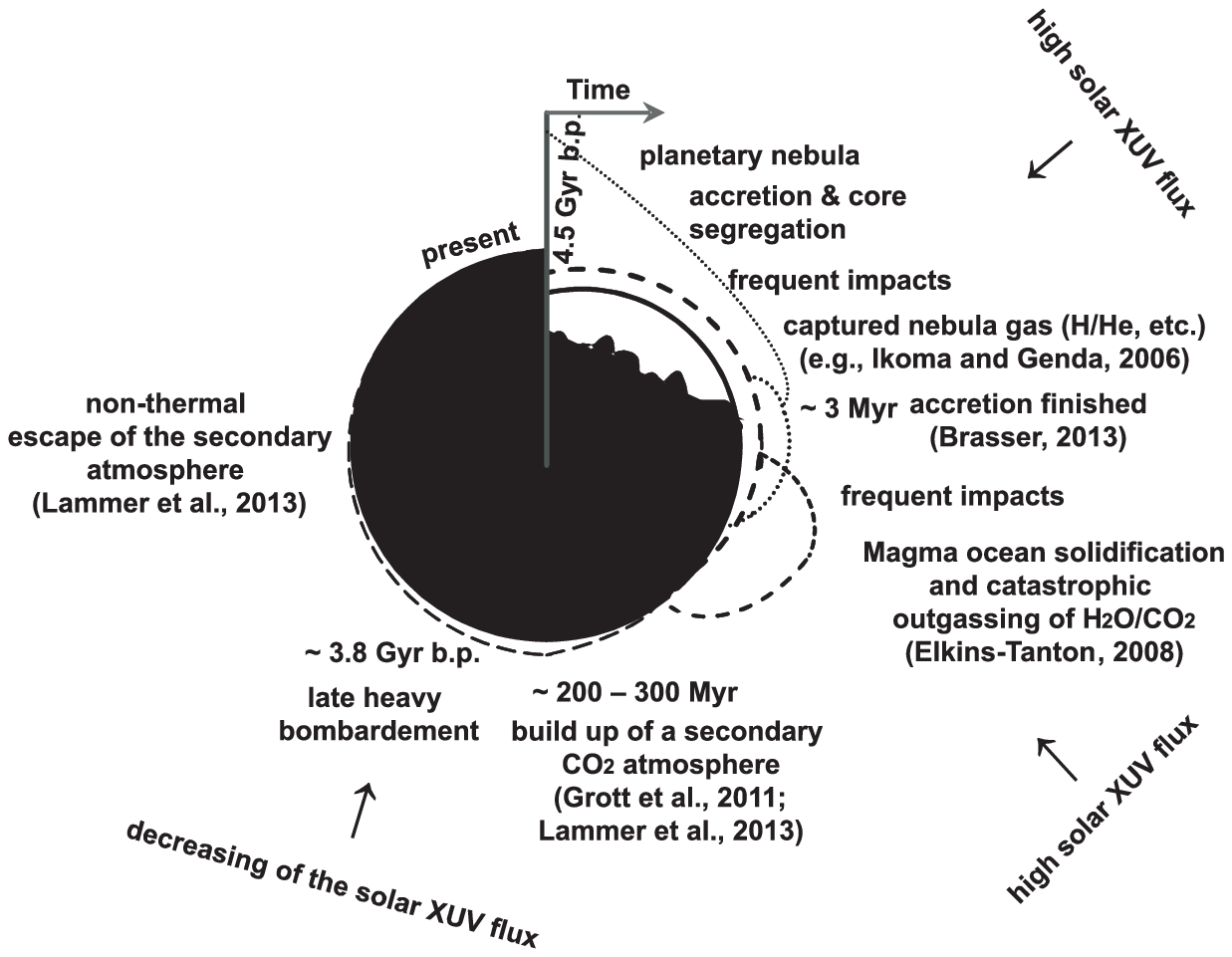}
\caption{Illustration of Mars' origin and protoatmosphere formation and evolution. The dotted lines
correspond to the accumulation during the growth and escape of nebula-based hydrogen from proto-Mars.
The onset of escape corresponds to the nebula dissipation time around $\sim$3-10 Myr, which is also
the expected time period when Mars finished its accretion (Brasser, 2012). The short dashed lines
illustrate the catastrophically outgassed volatiles and their expected escape after the planet's
magma ocean solidified.
Later on when the solar activity decreased a secondary CO$_2$ atmosphere could have build up by
volcanic activity (Grott et al., 2011; Lammer et al., 2013a) and the late heavy bombardment may also have delivered
volatiles to Mars $\sim3.8$ Gyr ago.}
\end{center}
\end{figure}
\begin{itemize}
\item capture of hydrogen and other gases (He, noble gases, etc.) from the solar nebular,
\item catastrophic outgassing of volatiles such as H$_2$O, CO$_2$, etc. and the formation of a steam atmosphere during and after the
magma ocean solidification period,
\item impact delivery of volatiles by asteroids and comets,
\item degassing by volcanic processes during geological epochs.
\end{itemize}
Fig. 1 illustrates the expected atmosphere formation and loss scenarios for Mars during the planet's history.
In the present work we focus on the origin and the evolution of the earliest martian protoatmosphere,
consisting of hydrogen accumulated from the solar nebular and a catastrophically outgassed steam atmosphere
after the planet finished its accretion and the magma ocean solidified.

\subsection{Captured hydrogen envelope around early Mars}
When proto-planets grow within the surrounding solar nebula by accretion of planetesimals, an extensive
amount of gas will be attracted so that optically thick, dense hydrogen envelopes accumulate around a rocky
core (e.g., Mizuno et al., 1978; Hayashi et al., 1979; Wuchterl, 1993; Ikoma et al., 2000; Ikoma and Genda, 2006;
Rafikov, 2006). The structure of such nebular-based hydrogen atmospheres was investigated decades ago
by Hayashi et al. (1979) and Nakazawa et al. (1985) for a wide range of planetary accretion rates,
grain opacities, and gas disk densities. These pioneering studies obtained captured nebula gas
around a Mars-mass body (i.e. $\sim 0.1M_{\rm \bigoplus}$) of $8.4\times 10^{24}$ g during
nebular life times of $\sim$1--10 Myr, equivalent to the
hydrogen content of $\sim$55 Earth oceans (1EO$_{\rm H}\approx 1.53 \times 10^{23}$ g).
More recent studies improved on these earlier results by adoption of
realistic gas and dust opacities as well as a realistic equation of state leading to
significantly lower atmosphere masses around bodies with masses that are $\sim 0.1M_{\rm \bigoplus}$
(Ikoma and Genda, 2006).

For the present investigation we computed a set of atmospheric models for Mars
to obtain an estimate of the amount of gas collected from the protoplanetary disk into the
planetary atmosphere. The hydrostatic structure equations have been solved by using the initial
model integrator of the adaptive, implicit RHD-Code (TAPIR-code) the equation of state from Saumon et al.~(1995),
gas opacities from Freedman et al.~(2008),
and dust opacities by Semenov et al.~(2003).
Convective energy transport is included in TAPIR in the form of a
turbulent convection model
loosely based on the description by Kuhfu{\ss} (1986).
For details on the parametrization and a short discussion of related convection models see
Freytag and St{\"o}kl (2013).

For the conditions of the solar nebula at the position of the Mars orbit we assumed
a gas density of $5\times 10^{-10}$ g cm$^{-3}$ and a temperature of 200 K.
These values are in good agreement with restrains derived from the minimum-mass solar nebula (Hayashi, 1981).
The minimum-mass solar nebula (MSN) is a protoplanetary disk that contains the minimum amount of solid material which is
necessary to build the planets of the Solar system.

The outer boundary conditions, i.e. nebula density and temperature,
have been implemented at the Hill radius $r_{\rm Hill}$ for all models
as we consider $r_{\rm Hill}$ to be a good approximation for the place
where the essentially hydrostatic structure of the planetary atmosphere blends into the background disk structure.
However, when calculating the captured atmospheric masses, i.e. the amount of gas in effect gravitationally bound to the planet,
we used the minimum of $r_{\rm Hill}$ and the Bondi radius $r_{\rm Bondi}$, which turns out to be equal to the
latter for all model runs by a margin of about a magnitude. The definition of the outer boundary condition seems to be,
apart from the equation of state and nebular opacities, the main cause for the different captured atmospheric masses obtained by
different authors. According to Ikoma (2012; private communication), the discrepancy between Ikoma and Genda (2006) and
Hayashi et al. (1978) is a case in point. In general, as also described by Ikoma and Genda (2006), the atmospheres (and thus the atmospheric masses)
of low-mass planets such as Mars are more dependent
on outer boundary conditions than atmospheres of more massive Earth-like and super-Earth-type cores.

In order to get some measure of the uncertainties involved in our modeling, we covered a small parameter space by varying
the most important atmospheric parameters: the planetary luminosity $L_{\rm pl}$ and the dust grain depletion factor $f$.
Table~\ref{atmoresults} and Fig.~2 summarize the results of our atmospheric calculations.

\begin{table*}
\label{atmoresults}
\caption{Integral parameters for Mars model atmospheres models with dust depletion factors $f$ of 0.1, 0.01, and 0.001 and for
accretion rates $\dot{M}_{\rm acc}$ between $1 \times 10^{-6}$ and $1 \times 10^{-9}$ Earth masses per year.
$L$ is the luminosity resulting from the accretion of planetesimals;
$M_{\rm atm}$ is the atmospheric mass up to the Bondi radius;
and surface pressure and temperature on the surface are denoted as $P_{\rm s}$ and $T_{\rm s}$, respectively.}
\begin{center}
\begin{tabular}{cccccc}
$\dot{M}_{\rm acc}$ [$M_{\rm Mars}$/yr] & $f_{\rm dust}$ & $L$ [erg/s]           & $M_{\rm atm}$ [g]     & $P_{\rm s}$ [bar]  & $T_{\rm s}$ [K] \\\hline
$9.35 \times 10^{-9}$                   & 0.001          & $2.39 \times 10^{22}$ & $6.58 \times 10^{22}$ & 7.81                  & 600                \\
$9.35 \times 10^{-9}$                   & 0.01           & $2.39 \times 10^{22}$ & $3.21 \times 10^{22}$ & 3.38                  & 639                \\
$9.35 \times 10^{-9}$                   & 0.1            & $2.39 \times 10^{22}$ & $1.00 \times 10^{22}$ & 8.08                  & 690                \\\hline
$9.35 \times 10^{-8}$                   & 0.001          & $2.38 \times 10^{23}$ & $2.66 \times 10^{22}$ & 2.58                  & 693                \\
$9.35 \times 10^{-8}$                   & 0.01           & $2.38 \times 10^{23}$ & $9.76 \times 10^{21}$ & 0.763                 & 724                \\
$9.35 \times 10^{-8}$                   & 0.1            & $2.38 \times 10^{23}$ & $2.86 \times 10^{21}$ & 0.155                 & 754                \\\hline
$9.35 \times 10^{-7}$                   & 0.001          & $2.38 \times 10^{24}$ & $8.81 \times 10^{21}$ & 0.628                 & 795                \\
$9.35 \times 10^{-7}$                   & 0.01           & $2.38 \times 10^{24}$ & $2.84 \times 10^{21}$ & 0.151                 & 784                \\
$9.35 \times 10^{-7}$                   & 0.1            & $2.38 \times 10^{24}$ & $5.25 \times 10^{20}$ & 0.028                 & 841                \\\hline
$9.35 \times 10^{-6}$                   & 0.001          & $2.38 \times 10^{25}$ & $2.70 \times 10^{21}$ & 0.132                 & 885                \\
$9.35 \times 10^{-6}$                   & 0.01           & $2.38 \times 10^{25}$ & $5.14 \times 10^{20}$ & 0.028                 & 862                \\
$9.35 \times 10^{-6}$                   & 0.1            & $2.38 \times 10^{25}$ & $3.21 \times 10^{19}$ & 0.005                 & 960                \\
\end{tabular}
\end{center}
\end{table*}
\begin{figure}
\includegraphics[width=0.94\columnwidth]{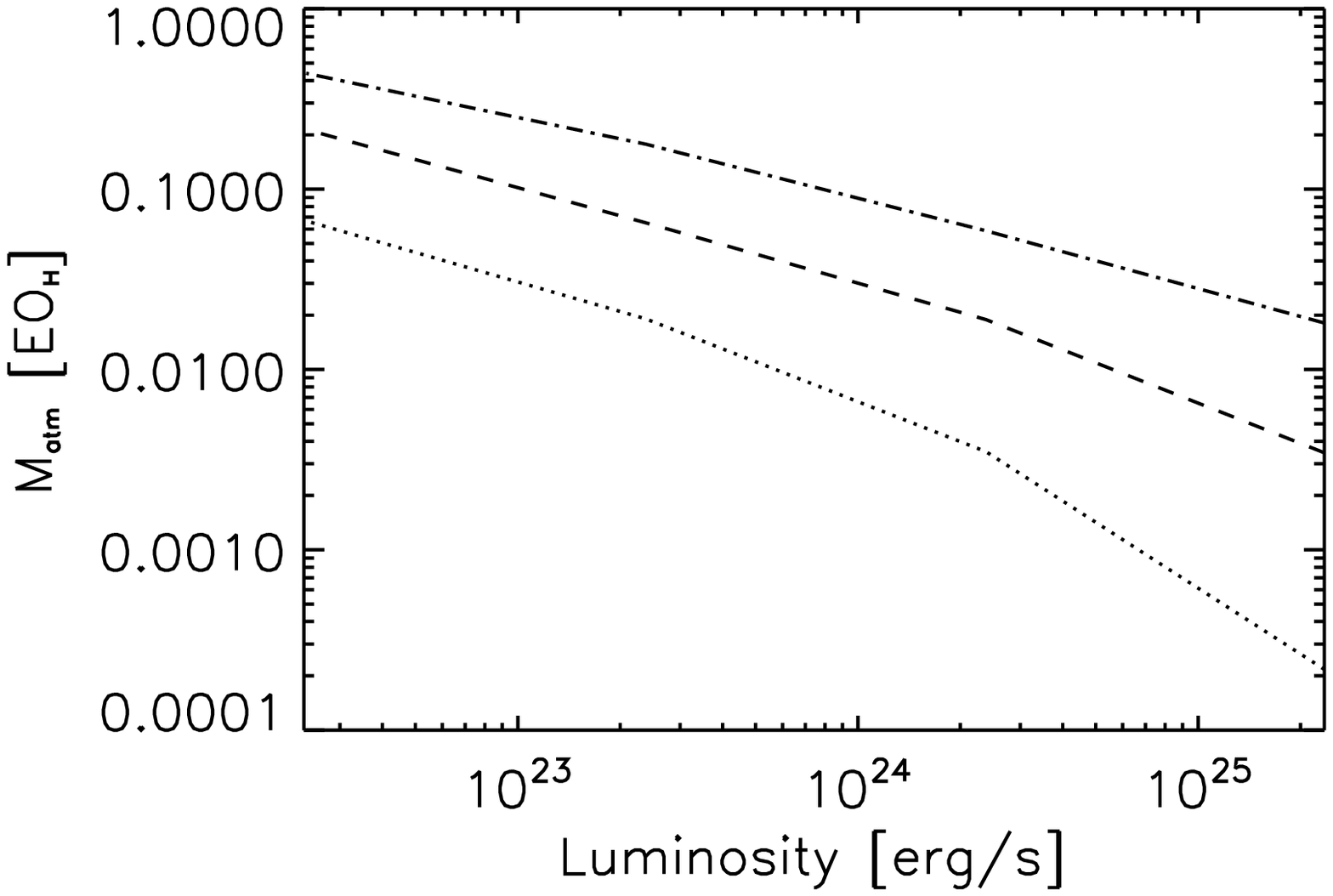}
\caption{Nebular-captured hydrogen envelopes for a Mars size and mass object at 1.5 AU, in units of Earth ocean equivalent amounts
of hydrogen (1EO$_{\rm H}$ = $1.53\times 10^{23}$ g) as a function of luminosity for three different dust grain depletion factors 
$f=0.001$ (dashed-dotted line), $f=0.01$ (dashed line), $f=0.1$ (dotted line).}
\end{figure}
$L_{\rm pl}$ is related to the rate of infalling planetesimals
\begin{equation}
L_{\rm pl} \simeq G M_{\rm pl} \dot{M}_{\rm acc} \left( \frac{1}{r_{\rm pl}} - \frac{1}{r_{\rm Hill}}\right),
\end{equation}
with $G$ the Newton gravitational constant, planetary mass $M_{\rm pl}$, planetary radius $r_{\rm pl}$
and planetesimal accretion rate $\dot{M}_{\rm acc}$.
Taking into account that according to Walsh et al.\ (2011) and Brasser (2013) Mars' formation was completed before
or soon after the nebular gas disappeared at $\sim$3--10 Myr, $M_{\rm pl}/\dot{M}_{\rm acc}$
should be several $\sim10^6$ years or larger.
On the other hand, according to Elkins-Tanton (2008) and Hamano et al. (2013) the cooling time scale of a Mars-size planet could be well above 1 Myr
and thus it seems plausible that during the nebula-gas accumulation phase the heat flux from the interior significantly adds to the planetary luminosity.
The lower limit of $L_{\rm pl}$ can be constrained by the radiogenic luminosity estimated
to be $\sim 10^{20}$ erg s$^{-1}$ for Mars (W\"{a}nke and Dreibus, 1988).

For higher planetary luminosities $T_{\rm s}$ almost reaches 1000 K and it is well likely that
models with, e.g. other boundary conditions or different dust opacity data, yield even higher surface temperatures.
It is important to note that H$_2$O can also be produced on a planet if $T_{\rm s}> 1500$ K. In such a case the planet's surface melts and
atmospheric hydrogen can be oxidized by oxides such as w\"{u}stite, magnetite and fayalite,
which are inside the planet to produce H$_2$O on the planet (Sasaki, 1990; Ikoma and Genda, 2006).
However, the model results which yield high surface temperatures are also those with only comparatively thin hydrogen envelopes,
which is reasonable as high luminosities and temperatures tend to inflate a planetary atmosphere.
Therefore, one may speculate that Mars atmospheres with $T_{\rm s}> 1500$ K will be too thin to allow for
efficient H$_2$O production from a captured and oxidized hydrogen envelope.

Before we discuss the radiation environment of the young Sun
during the first 100 Myr after Mars' origin and before we model the escape of the nebula-based hydrogen envelope
we investigate the possible range of catastrophically outgassed steam atmospheres.
\subsection{Magma ocean and outgassing of a steam atmosphere on early Mars}
As discussed before the terrestrial planets are thought to have reached their final sizes by a series of giant accretionary impacts.
These impacts were energetic enough to produce melting of some depth in the planet (e.g., Tonks and Melosh, 1993; Reese and Solomatov, 2006;
Lebrun et al., 2013).
This hypothesis is supported by the discovery of $^{142}$Nd isotope anomalies in martian SNC
meteoroids, which indicate that early Mars developed a magma ocean
(Harper et al., 1995; Foley et al., 2005; Debaille et al., 2007). Therefore, the first major degassed volatile-rich atmospheres
likely resulted from the solidification of these magma bodies, and their release into the growing atmosphere
in excess of what can be held in crystallizing silicate minerals
(Abe, 1993; 1997; Abe and Matsui, 1988; Matsui and Abe, 1986; Zahnle et al., 1988; Elkins-Tanton et al., 2005; Debaille et al., 2007; Elkins-Tanton, 2008; 2011;
Hamano et al., 2013; Lebrun et al., 2013).
In these models the magma ocean is expected to solidify from the bottom upward, because the slope of the adiabat is steeper than the
slope of the solidus and thus they first intersect at depth. Because the energy and size of late accretionary impacts on early Mars are
unknown, we consider a 500 km-deep magma ocean and, as an end-member, a 2000 km-deep or whole mantle magma ocean.

\begin{table*}
\caption{Modelled atmospheric partial surface pressures $P_{\rm H_2O}$ and $P_{\rm CO_2}$ in units of bar of catastrophically
outgassed steam atmospheres dependent on initial H$_2$O and CO$_2$ contents in wt\% inside a magma ocean with a minimum depth
of 500 km and a maximum depth of 2000 km.}
\begin{center}
\begin{tabular}{l|cc|cc}
Bulk magma ocean & initial H$_2$O [wt.\%]& initial CO$_2$ [wt.\%] & $P_{\rm H_2O}$ [bar] & $P_{\rm CO_2}$ [bar] \\\hline
500 km deep      &                      &                                                             \\\hline
                 &   0.1                & 0.02                  & 52                  &   11          \\
                 &   0.2                & 0.04                  & 108                 &   22          \\\hline
2000 km deep     &                      &                       &                     &               \\\hline
                 &   0.1                & 0.02                  & 122                 &   26           \\
                 &   0.2                & 0.04                  & 257                 &   54           \\
\end{tabular}
\end{center}
\end{table*}
H$_2$O and CO$_2$ will be integrated in solidifying minerals in small quantities, will be enriched in solution in magma ocean liquids as
solidification proceeds, and will degas into a growing steam atmosphere. At pressures and temperatures of magma ocean crystallization
no hydrous or carbonate minerals will crystallize (Ohtani et al., 2004; Wyllie and Ryabchikov, 2000). Details of the solidification process,
the mineral considered, their H$_2$O and carbon partitioning, and other methods can be found in Elkins-Tanton (2008).
The quantity of water and carbon compounds available for degassing is dependent upon the bulk composition of the magma ocean.
The terrestrial planets are likely to have been accreted from chondritic material and planetesimals built from chondrites.

Alexander et al. (2012) recently demonstrated that Earth's water, and therefore likely Mars' water, originated mainly from rocky meteoritic material.
Wood (2005) reports up to 20 wt\% of H$_2$O in primitive undifferentiated chondrites, and Jarosewich (1990) reports $\sim$3 wt\% H$_2$O
in achondrites, though most are drier. Enstatite chondrites match the oxygen
isotope composition of the Earth, but smaller fractions of the wide compositional range of other meteorite compositions
(see also Alexander et al., 2012 and Drake and Righter, 2002; and references therein)
though volatile-rich material from greater radii in the planetary disk may have been added later
in planetary formation (e.g., Raymond et al., 2006; O'Brien et al, 2006).
Here we assume water and carbon is added to the growing rocky planets from rocky chondritic material.

Though the original quantity of water and carbon added during giant impacts remains unconstrained, we model
two possible starting compositions, according to Brasser (2013) one with 1000 ppm H$_2$O, and one with 2000 ppm H$_2$O,
each with one-fifth the CO$_2$ content. These initial compositions are conservatively supported by the data of Jarosewich (1990).
For simplicity the carbon is assumed to be degassed as CO$_2$, though reducing conditions may have produced CO or even CH$_4$.

Elkins-Tanton (2008) showed that for a range of magma ocean bulk compositions with between $\sim$500--5000 ppm H$_2$O, between $\sim$70\% and $\sim$99\%
of the initial water and carbon is degassed into the planetary atmosphere. Magma ocean solidification is therefore the most significant
degassing event in a planet's evolution; the remainder of the volatiles are stored in the interior,
available for later degassing via volcanic processes (e.g., Grott et al., 2011).

Table 2 shows the partial surface pressures of catastrophically outgassed steam atmospheres,
depending on the assumed bulk magma ocean depths
and the initial H$_2$O and CO$_2$ contents in the magma ocean in wt.\% according to the model of Elkins-Tanton (2008).
One can see that a global magma ocean with the depth of $\sim$500 km can produce
a steam atmosphere with total surface pressures of $\sim$60--130 bar. If the
magma ocean contained the whole mantle, surface pressures between $\sim$150--310 bar could have been outgassed.
\section{Radiation environment during Mars' initial life time}
The efficiency of thermal atmospheric escape is related to the planet's temperature at the
base of the thermosphere which is located near the mesopause-homopause location in combination with
the amount of the XUV flux that is absorbed in the upper atmosphere.
The predicted evolution of the Sun's bolometric luminosity relative to its present value and the related equilibrium
temperature $T_{\rm eq}$ at Mars in shown in Fig.~3. We have chosen two stellar evolution tracks (Baraffe et al., 1998; Tognelli et al., 2011)
which predict the lowest and highest luminosities, respectively, between 1 and 10~Myr compared to other authors
(cf. Fig.~14 of Tognelli et al., 2011). From Baraffe et al. (1998), the track with parameters $M=1M_\odot$, $Y=0.282$, $Z=0.02$
and mixing length parameter $\alpha=1.9$ was adopted, the track from Tognelli et al. (2011) has $M=1M_\odot$, $Y=0.288$,
$Z=0.02$ and $\alpha=1.68$.
\begin{figure}
\includegraphics[width=0.85\columnwidth]{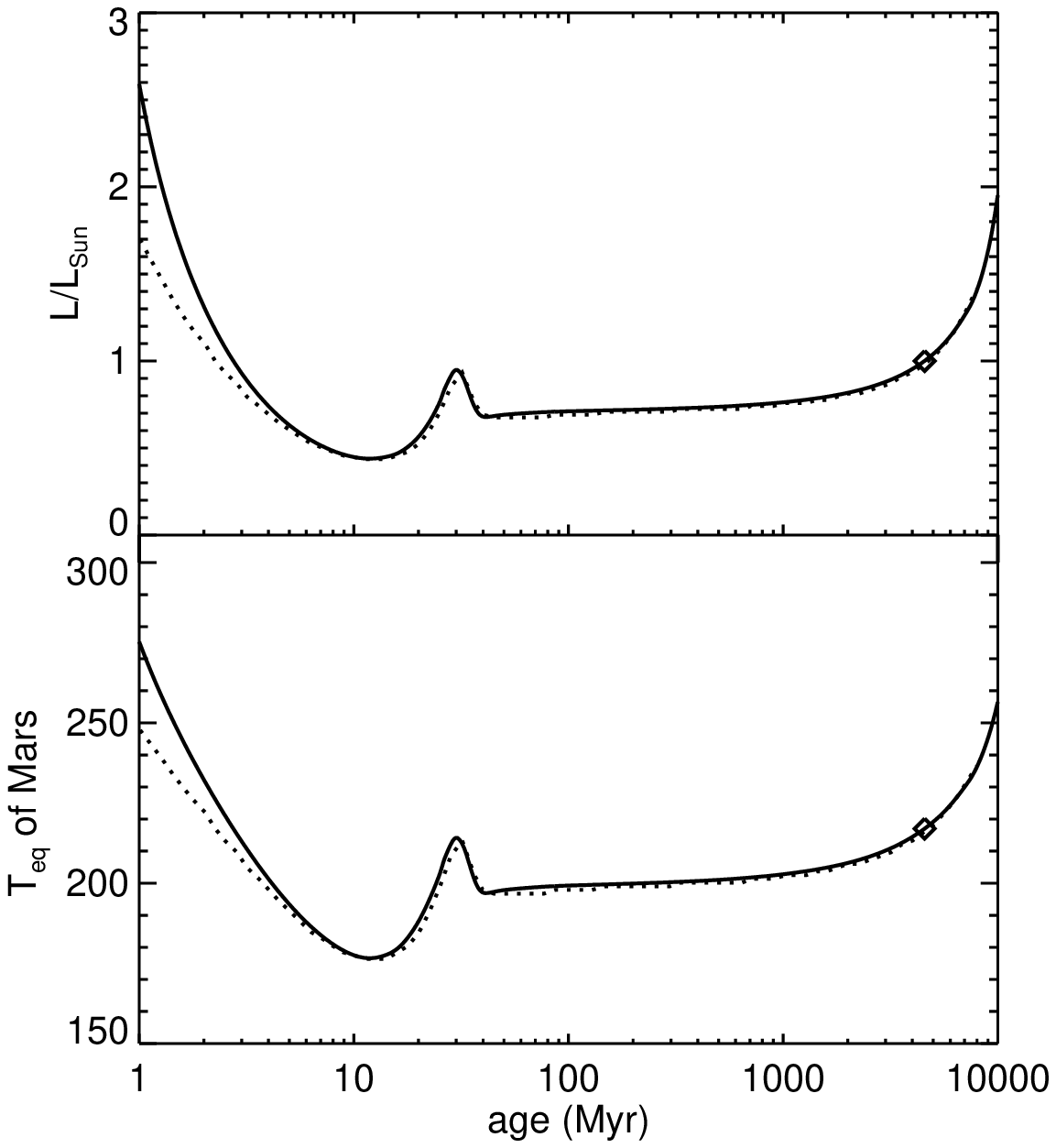}
\caption{Evolution of the Sun's bolometric luminosity relative to its present value (upper panel)
and the equilibrium temperature of Mars (lower panel). The solid line corresponds to an evolution
track of Tognelli et al. (2011) and the dotted line to Baraffe et al. (1998), both for a star of solar mass and
metallicity. The present-day values in both panels are indicated by diamonds. For the evolution of the $T_\mathrm{eq}$,
a constant albedo was assumed.}
\end{figure}
For planetary atmospheres that are in long-term radiative equilibrium the so-called planetary skin temperature is $T_{\rm eff}\approx T_{\rm eq}$.
The lower panel of Fig. 3 shows the corresponding evolution of the equilibrium temperature of Mars,
which is about 200 K, 3--4 Myr after the Sun's origin. We assume a constant albedo over time and adopt a
present-day value of $T_{\rm eq}=217$~K. One should note that variations of the albedo due the evolution of Mars' early atmospheric composition
and the Sun's spectral energy distribution could alter the predicted
evolution of $T_{\rm eq}$ shown in Fig. 3.
Thermal escape of
the martian protoatmosphere
was driven by the XUV emission of the young Sun. The
evolution of this high-energy emission of a solar-type star can
be roughly divided into two regimes, the saturation phase and
the post-saturation evolution. During the saturation phase the
stellar X-ray flux does not scale with the stellar rotation
period and is saturated about 0.1\% of the bolometric
luminosity $L_\mathrm{bol}$ (Pizzolato et al., 2003; Jackson et
al., 2012).
After the Sun settled on the main sequence and began to
spin down from a possibly shorter period to about 2 days
due to spin-down via angular momentum loss by the
solar wind, the post-saturation phase began. During this phase,
the XUV emission of the Sun was determined by its rotation
period. A reconstruction of the XUV-evolution during this time
period was attempted in the ``Sun in Time'' program (G\"udel,
2007 and references therein). By studying a sample of solar
analogs of different ages Ribas et al. (2005) found that the
Sun's XUV flux enhancement factor $I_\mathrm{XUV}$ at Earth's orbit in the
wavelength range 1--1200~\AA\ can be calculated as
\begin{equation}
I_\mathrm{XUV}=\left(t/4.56\right)^{-1.23}
\end{equation}
with the age $t$ in Gyr. This relation was calibrated back to
an age of 100~Myr corresponding to the youngest solar analog in
their sample. However, deviations from this power law are
possible during the first few hundred Myr because the stellar
rotation periods, which determine the efficiency of the
magnetic dynamo and, hence, the XUV emission during this phase,
are not unique.

The XUV-evolution during the saturation phase was, as mentioned
above, mainly determined by the evolution of $L_\mathrm{bol}$.
Due to the difficulty of observing stars in the EUV because of
the strong absorption by the interstellar medium, much of what
is known about the high-energy emission of very young stars is
extrapolated from X-ray observations. Between the zero-age main
sequence (ZAMS), which the Sun reached at an age of about
50~Myr according to stellar evolution models (e.g. Baraffe et
al., 1998; Siess et al., 2000), and the end of the saturation
phase, the solar XUV flux should have been approximately
constant because of the more or less constant bolometric
luminosity. For pre-main sequence (PMS) stars, the observed
X-ray luminosities are in the order of a few
$10^{30}\,\mathrm{erg\,s^{-1}}$ and show a large spread of more
than an order of magnitude (Preibisch et al., 2005; Telleschi et al., 2007). These
values are nevertheless consistent with the saturation level of main-sequence stars
mentioned above because of the more luminous PMS-Sun and the
observed evolution of the stellar X-ray emission during the
first tens of Myr seems to be determined mainly by changes of
$L_\mathrm{bol}$ (Preibisch et al., 2005; Briggs et al., 2007).

The estimated past evolution of the Sun's XUV flux, scaled to
the orbit of Mars and normalized to the average present solar
value of $2\,\mathrm{erg\,cm^{-2}\,s^{-1}}$ (scaled from the
present value at Earth of
$4.64\,\mathrm{erg\,cm^{-2}\,s^{-1}}$; Ribas et al., 2005), is
shown in Fig.~4. The solid line indicates the post-saturation
evolution after Eq.~2 and the symbols correspond to data from
solar analog stars and the Sun. The dotted lines sketch a
possible PMS-XUV evolution based on the evolution of
$L_\mathrm{bol}$ using theoretical evolutionary tracks for a
solar mass star (Baraffe et al., 1998) and assuming that the XUV
emission consists mainly of X-rays, so that
$L_\mathrm{XUV}/L_\mathrm{bol} \approx
L_\mathrm{X}/L_\mathrm{bol} \approx 10^{-3.2\pm0.3}$. The value
of the saturation level is adopted from Pizzolato et al. (2003)
for stars of about one solar mass. The uncertainties of the
Sun's XUV emission before the ZAMS are large because of the dependence of its activity level on the convection
zone depth and the rotational history, which in turn depends on the disk locking history.
Moreover, the contribution of EUV to the total XUV flux is observationally unconstrained because of strong
absorption by the interstellar medium. Therefore we adopt a
constant average XUV flux level of about 100 times the present value for our escape rate calculations.

The shaded area indicates the approximate formation time of
Mars which occurred during the first few Myr (Brasser, 2013).
The inner disk was still present after Mars formed, and the
inner planets were still forming. An inner disk would have
absorbed a significant fraction of the Sun's XUV radiation
until it became optically thin so that the XUV flux actually
received by Mars could have been lower than estimated in
Fig.~4. Typically, inner disks disperse on timescales within
a few Myr to 10 Myr (e.g. Mamajek et al., 2004; Najita et
al., 2007; Hillenbrand, 2008).
\begin{figure}[h]
\begin{center}
\includegraphics[width=0.85\columnwidth]{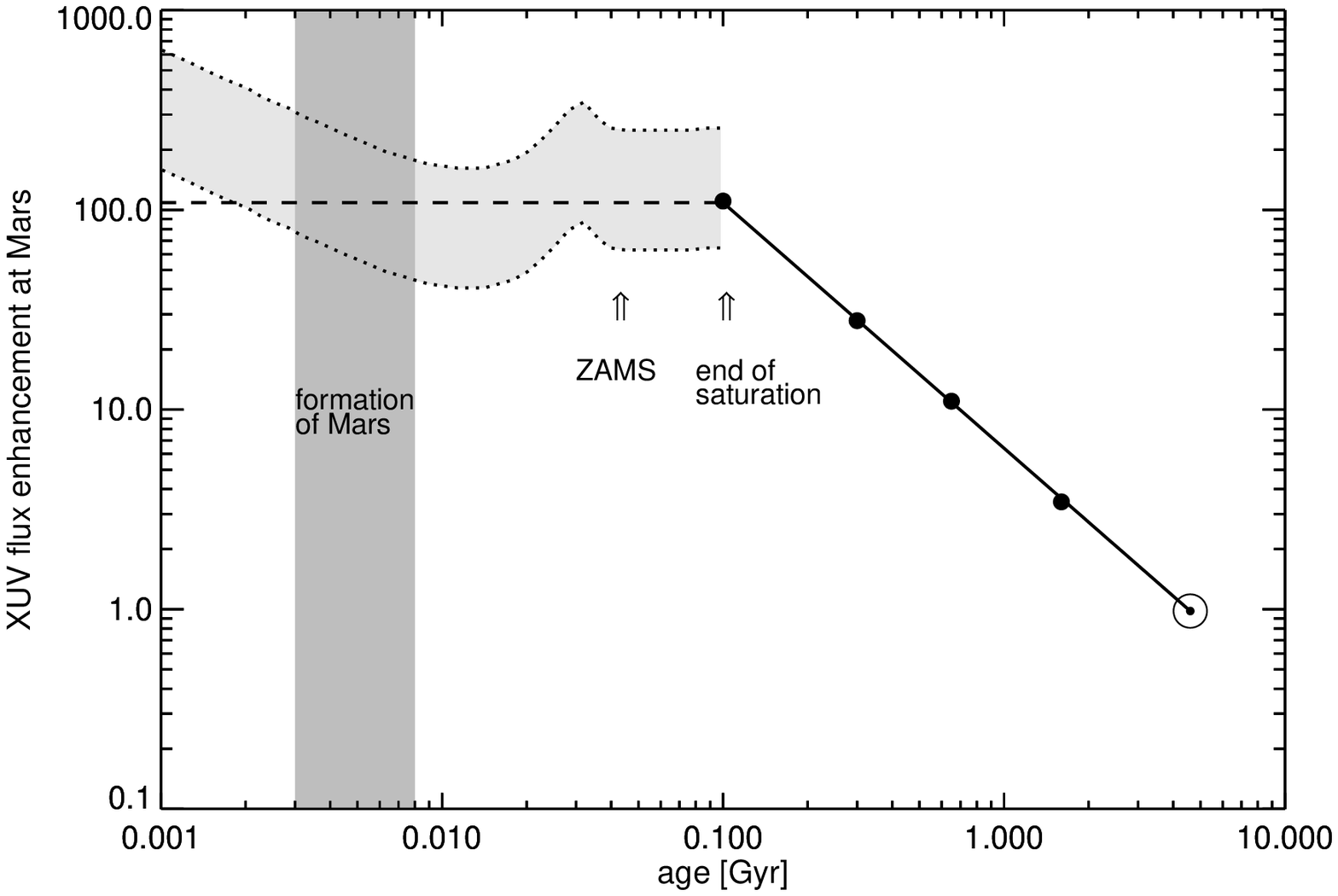}
\caption{Evolution of the Sun's XUV emission normalized to the present value and scaled to the present martian orbit at 1.52 AU.
The solid line indicates the evolution during the post-saturation phase (Ribas et al. 2005) with data of solar analogs (black dots)
and the Sun indicated. The dotted lines indicate the approximate evolution of the saturated XUV emission estimated by
$10^{-3.2\pm0.3}L_\mathrm{bol}$ (Pizzolato et al. 2003), with the bolometric luminosity taken from stellar evolution tracks of
a solar mass star (Baraffe et al. 1998). The shaded area indicates the expected formation time of Mars (Brasser 2013). The
dashed line shows our adopted average XUV value during the Sun's saturation phase.}
\end{center}
\end{figure}
Thus, if one compares the latest views of Mars' origin and age with that of the
radiation history of the young Sun and the nebula dissipation time, Mars' nebula-based
and/or outgassed steam atmosphere as well as volatiles which were delivered by
frequent impacts were exposed to an XUV flux which was $\sim$100 times stronger compared to that
of the present Sun during $\sim$95--100 Myr after the planet's origin. In the following section
we investigate how long early Mars could have kept these hydrogen-rich protoatmospheres against XUV-driven
thermal atmospheric escape.
\section{Thermal escape of Mars' protoatmosphere}
At present Mars the CO$_2$-rich thermosphere is in hydrostatic equilibrium, while
a hydrogen-rich upper atmosphere of the protoatmosphere that is exposed to the high
XUV flux of the young Sun will hydrodynamically expand and the bulk atmospheric particles
can escape efficiently (e.g., Watson et al., 1981; Chassefi\`{e}re, 1996a; 1996b; Tian et al., 2009; Lammer, 2013; Lammer et al., 2012;
2013a). For this reason we apply a 1-D hydrodynamic upper atmosphere model to the martian protoatmosphere
and calculate the XUV-heated hydrogen-dominated dynamically expanding upper atmosphere structure and
the thermal hydrogen escape rates, including dissociated and
dragged heavier atmospheric main species.
\subsection{Energy absorption and model description}
The thermosphere is heated due to the absorption, excitation, dissociation and ionization of
the gas by the incoming solar XUV radiation. By averaging the XUV volume heating rate over Mars' dayside
the volume heat production rate $q_{\rm XUV}$ due to the absorption of
the solar radiation can then be written as (e.g., Erkaev et al., 2013;
Lammer et al., 2013b)
\begin{equation}
q(t,r)=\frac{\eta n \sigma_{\rm a}}{2}\int_{0}^{\frac{\pi}{2}+\arccos(\frac{1}{r})} J(t,r,\Theta)
\sin\Theta d\Theta,
\end{equation}
with the polar angle $\Theta$ and $J(t,r,\Theta)=J_{\rm
XUV}e^{-\tau(t,r,\Theta)}$, where
\begin{eqnarray}
\tau(t,r,\Theta) = \int_{r\cos \Theta}^\infty {\sigma_{\rm a} n\left(t,\sqrt{s^2+r^2 \sin^2\Theta}\right) \,d s},
\end{eqnarray}
$q$ is the volume heating rate depending on the radial distance, $n$ the atmospheric number
density which is a function of time and spherical radius $r$,
$\eta$ the heating efficiency which corresponds to the fraction
of absorbed XUV radiation which is transformed into thermal
energy. Depending on the availability of IR-cooling molecules
such as H$_3^+$ or CO$_2$ it is known from various studies that
$\eta\sim$15--60 \% (Chassefi\`{e}re, 1996a; 1996b; Yelle, 2004;
Lammer et al., 2009; Leitzinger et al., 2011; Koskinen et al.,
2013). $\sigma_{\rm a}$ is the absorption cross-section of
hydrogen, and $J_{\rm XUV}$ is the XUV energy flux of the young
Sun outside the protoatmosphere.

For studying the XUV-exposed structure of the upper atmosphere we solve the system of the 1-D fluid
equations for mass, momentum, and energy conservation in spherical coordinates by applying a non-stationary
1D hydrodynamic upper atmosphere model which is described in detail in
Erkaev et al. (2013)
\begin{equation}
\frac{\partial \rho r^2}{\partial t} + \frac{\partial \rho v
r^2}{\partial r}= 0,
\end{equation}
\begin{eqnarray}
\frac{\partial \rho v r^2}{\partial t} + \frac{\partial \left[ r^2 (\rho v^2+P)\right]}{\partial r} =\rho g r^2 + 2P r,\\
\frac{\partial r^2\left[\frac{\rho v^2}{2}+\frac{P}{(\gamma-1)}\right]}{\partial t}
+\frac{\partial v r^2\left[\frac{\rho v^2}{2}+\frac{\gamma P}{(\gamma - 1)}\right]}{\partial r}=\nonumber\\
\rho v r^2 g + q_{\rm XUV} r^2,
\end{eqnarray}
with pressure
\begin{equation}
P=\frac{\rho}{m_{\rm H}} k T,
\end{equation}
and gravitational acceleration,
\begin{equation}
g=-\nabla\Phi,
\end{equation}
We note that we neglect the conduction term in the equations because as shown later
the energy flux related to thermal conductivity is less important under these extreme conditions compared to the
energy flux of the hydrodynamic flow. Here, $\rho$, $v$, $P$ and $T$ are
the mass density, radial velocity, pressure and temperature of the atmosphere, $r$
is the radial distance from the center of the planet,
$m_{\rm H}$ is the mass of atomic hydrogen, $G$ is Newton's gravitational constant,
$\gamma$ is the polytropic index or the ratio of the specific heats,
and $k$ is the Boltzmann constant.

For computational convenience we introduce normalized parameters
\begin{eqnarray}
\tilde P = P /(n_0 k T_0), \quad \tilde\rho = \rho / (n_0 m), \nonumber \\
 \tilde v = v / v_0, \quad v_0 = \sqrt{k T_0 / m} , \quad
\tilde T = T/ T_0, \nonumber \\
\tilde q = q r_0 /(m n_0 v_0^3), \quad \tilde r = r/ r_0, \nonumber \\
\tilde t = t v_0/r_0, \quad  \beta = G m M_{\rm pl} /(r_0 k T_0).      \label{norm}
\end{eqnarray}
Here $r_0$, $T_0$, $n_0$ and $v_0$ are the radius, temperature,
number density and thermal velocity at the lower boundary of the
simulation domain. $\beta$ is the so-called
Jeans parameter (Chamberlain, 1963). For values of $\beta > 30$ the atmosphere
can be considered as bound to the planet. For values which are lower
classical Jeans escape happens. For $\beta$ values that are $\sim$2--3.5
the thermal escape can be very high (Volkov and Johnson, 2013) and for
for values $\leq$ 1.5 classical blow-off occurs and the
atmosphere escapes uncontrolled.
Using normalizations (\ref{norm}), we obtain the normalized XUV
flux distribution in the planetary atmosphere
\begin{eqnarray}
\tilde J(\tilde r,\Theta) = J / J_{XUV0}= exp[-\tilde\tau(\tilde r,\Theta)],
\end{eqnarray}
where
\begin{eqnarray}
\tilde\tau(\tilde r,\Theta) = \int_{\tilde r\cos \Theta}^{\infty}{a \tilde n\left(\tilde t, \sqrt{s^2 + \tilde r^2 \sin^2 \Theta} \right)d s},
\end{eqnarray}
where $a=\sigma_{\rm a} n_0 r_0$ is obtained due to the normalization of eq. (4). The normalized heating rate is given by
\begin{eqnarray}
\tilde q(\tilde r) = A \tilde n \int_0^{\pi/2+\arccos(1/\tilde r)}{\exp[-\tilde\tau (\tilde r,\Theta)]\sin \Theta d\Theta}, \label{q_norm}
\end{eqnarray}
Integrating (\ref{q_norm}) over the whole domain we obtain the
total energy absorption in the normalized units which is
proportional to the incoming XUV flux.
\begin{eqnarray}
\int_1^\infty{\tilde q 4\pi \tilde r^2 d\tilde r} = \pi\frac{J_{\rm XUV}}{m n_0 v_0^3}  \frac{r_{\rm XUV_{\rm eff}}^2}{r_0^2}, \label{Energy_balance}
\end{eqnarray}
where $r_{\rm XUV_{\rm eff}}$ is the effective radius of the XUV energy
absorption which is dependent on the density distribution. This
effective radius can be determined from the following equation
\begin{eqnarray}
r_{\rm XUV_{\rm eff}}^2/r_0^2 =1+2\int_1^\infty{[1-\tilde J(s,\pi/2)] s ds} .
\end{eqnarray}
As shown by Watson et al. (1981) the effective radius can exceed the planetary radius quite
substantially for a planetary body, which has a low gravity field and hence in low values of
the $\beta$ parameter when its atmosphere is exposed by high XUV fluxes.
We get the appropriate coefficient
\begin{equation}
A=\frac{\eta \sigma_{\rm a} r_0 J_{\rm XUV}}{2 m v_0^3}
\end{equation}
in formula (\ref{q_norm}) to satisfy eq. (\ref{Energy_balance}) for a
given value of $J_{\rm XUV}$.

\subsection{Boundary conditions at the lower thermosphere}
The boundary conditions at the lower boundary of our simulation domain are the gas temperature $T_0$, number density $n_0$ and the corresponding thermal velocity $v_0$ near the mesopause-homopause level $r_0$, that is at present martian conditions located near the base of the thermosphere.
The value of the number density $n_0$ at the base of the thermosphere can never be arbitrarily increased or decreased as much as by an order of magnitude, even if the surface pressure on a planet varies during its life time by many orders of magnitude. The reason for this is that the value of $n_0$ is strictly determined by the XUV absorption optical depth of the thermosphere.
The temperature $T_0$ at the base of the thermosphere $z_0=(r_0-r_{\rm pl})$ is determined only by the variation of the equilibrium or skin temperature of a planet, to which the base temperature $T_0$ is usually quite close. In a hotter environment corresponding to the catastrophically outgassed steam
atmosphere, which is for instance strongly heated by frequent impacts, $z_0$ and the above estimated XUV
effective radius $r_{\rm XUV}$ simply rises to a higher altitude where the base pressure retains the same constant value as in a less dense atmosphere.

Marcq (2012) studied with a 1-D radiative-convective atmospheric model the coupling between magma oceans
and outgassed steam atmospheres and found that for surface temperatures $T_{\rm s}$ $\geq$ 2350 K,
the radiative temperature
of a planetary atmosphere $T_{\rm eff}$ can rise from $\sim$230 K to $\sim$300--400 K,
while $T_{\rm eq}$ remains close to $\sim$200 K.
However, such extreme surface temperatures are only be reached
during the totally and partially molten stage of the magma ocean, which last only
for $\approx $0.1 Myr (Lebrun et al., 2013).
For this reason we assume in the following thermal escape calculations similar as in Fig. 3
a temperature $T_0$ of 200 K at the base of the thermosphere
which corresponds to the equilibrium $T_{eq}$, or skin temperature of Mars' orbit.
We point out that an uncertainty of $\pm 20$ K as shown in the evolutionary path of
$T_{\rm eq}$ in Fig. 3 does not have a
big influence in the modeled escape rates.
We assume an atomic hydrogen density $n_0=10^{13}$ cm$^{-3}$ at the lower boundary of the hydrogen-rich
protoatmosphere (e.g., Kasting and Pollack, 1983; Tian et al., 2005). According to Kasting and Pollack (1983),
similar number density values can be expected also to H$_2$O mixing ratios $\geq$50 \% in a humid steam-like
terrestrial planetary atmosphere.

The upper boundary of our simulation domain is chosen at
70$r_{\rm pl}$, but the results of our hydrodynamic model are
considered as accurate only until the Knudsen number $Kn$, which
is the ratio between the mean free path and the scale height, reaches 0.1 (Johnson et al., 2013).
Because of the high XUV flux the whole bulk atmosphere reaches the martian escape velocity below
or at this altitude level.

The high XUV flux of the young Sun will
dissociate most H$_2$ and H$_2$O molecules in the thermosphere so that the upper part of the studied protoatmospheres should be
mainly dominated by hydrogen atoms (Kasting and Pollack, 1983;
Chassefi\`{e}re, 1996a; Yelle, 2004; Koskinen et al., 2010; Lammer, 2013).
As it was shown by Marcq (2012), during periods of magma ocean related hot surface temperatures the tropopause
location in an overlaying steam atmosphere can move at
an Earth or Venus-like planet from its present altitude of $\sim$30--40 km up to higher altitudes of $\sim$300--550 km.
Depending on the
surface temperature and pressure of the steam atmosphere in such an environment the mesopause level
would then also move to higher altitudes.
By applying the model of Marcq (2012) to the outgassed steam atmospheres given in Table 2, we obtain
mesopause altitudes of $\sim$330--350 km, $\sim$450--465 km, $\sim$610--630 km and $\sim$750--850 km
for surface temperatures of $\sim$1500 K, $\sim$2000 K, $\sim$2500 K and $\sim$3000 K, respectively.
This mesopause altitudes have been estimated by detailed modeling of the lowest 600 km
of the steam atmospheres. The altitudes above 600 km are obtained from an extrapolation with a precision
of $\sim$20 km in the 600--700 km range and $\sim$50 km above 700 km.
The simulations used a grey approximation for the radiative transfer which can influence the
profile by setting the mesospheric temperature and thus scale height to a slightly different value,
but we don't expect this uncertainty changes these altitudes by more than 20 km.
As one can see, even in the most extreme case with a surface temperature of 3000 K,
the mesopause altitude lies below 1000 km for a body with Earth's gravity. However,
it will most likely be higher than 1000 km with a lower gravity such as
Mars`. We plan to study the response to the mesopause location and its influence in the
escape of outgassed steam atmospheres on Mars in detail in the near future.

However, for illustrating the importance and influence of the mesopause location in the escape efficiency
we modeled also a case where we assumed that $z_0$ is located at 1000 km
above the planet's surface. That hydrogen-dominated gas envelopes
with hot surface temperatures will have larger radii compared
to planets with present time atmospheres is also addressed in Mordasini et al. (2012).
However, the planetary mass-radius relationship model results
for small and low mass bodies remain highly uncertain.
\section{Results}
\subsection{Thermospheric profiles and escape rates}
By exposing the martian protoatmospheres with a 100 times higher XUV flux compared to today's solar
value in martian orbit, we find that the convective thermal energy flux is less significant than the
thermal energy flux related to the hydrodynamic flow. Fig. 5 compares the thermal energy flux due
to the hydrodynamic flow (curves at the top: dotted lines: $\eta$=15\%;
dashed-lines: $\eta$=40\%) per steradian of the atmospheric particles with
the convective thermal energy flux (curves at the bottom: dotted lines: $\eta$=15\%;
dashed-lines: $\eta$=40\%), obtained by our hydrodynamic model.
The two sudden decreases in the convective thermal energy flux curves
can be explained, because this flux is proportional
to the temperature gradient, and therefore it decreases in the vicinity of the
temperature maximum and minimum. At first point we have a strong temperature maximum,
and at the second point we have shallow temperature minimum. By comparing the two
fluxes one can conclude that under such extreme conditions
the influence of the thermal conduction on the atmospheric escape is expected
to be rather small. Therefore we neglect the thermal conduction term in the energy equation.
\begin{figure*}[!ht]
\begin{center}
\includegraphics[width=0.9\columnwidth]{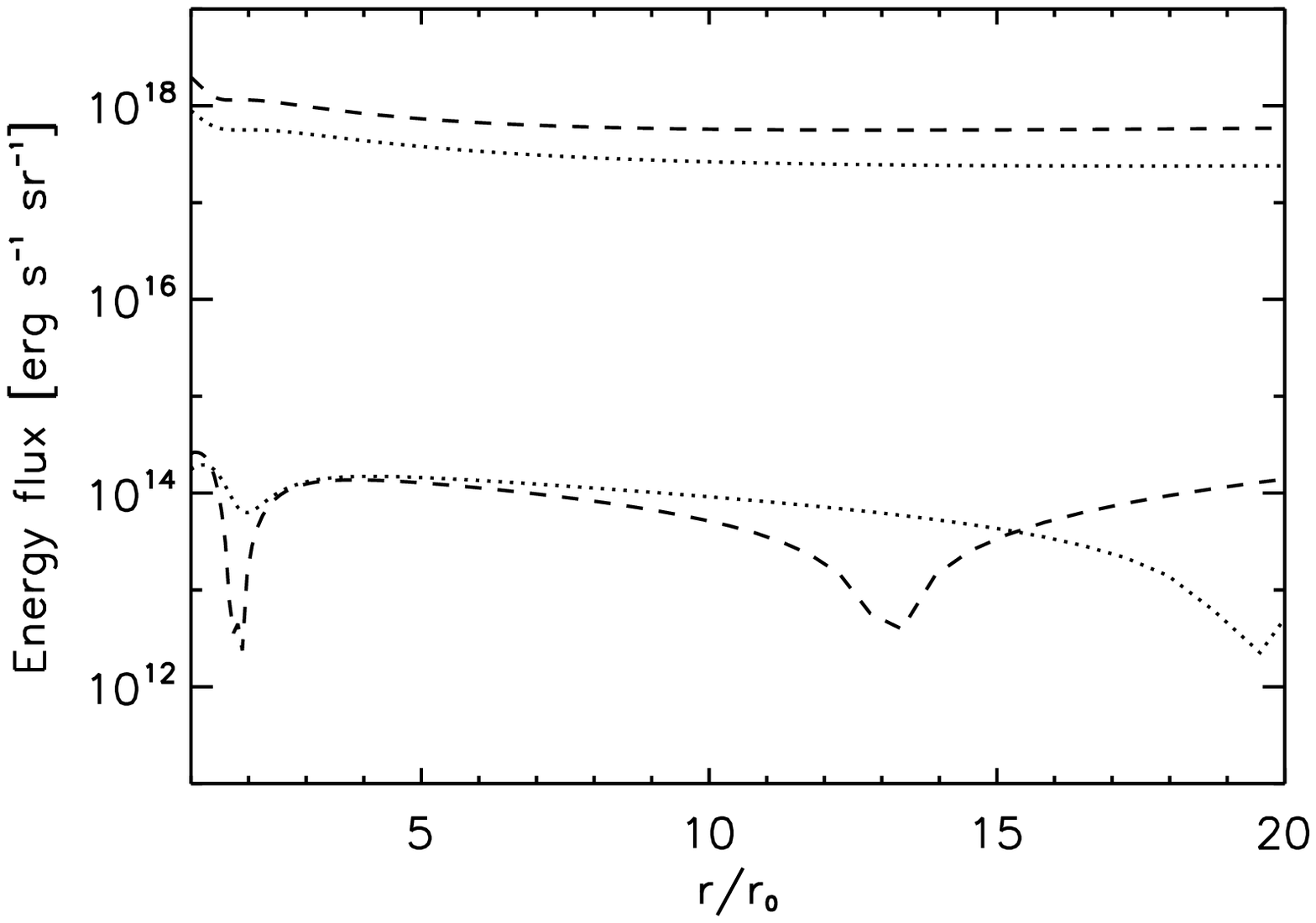}
\caption{Comparison of the thermal energy flux per steradian of the hydrodynamical flow (upper dashed
line: $\eta$=40\%; dotted line: $\eta$=15\%) with
the thermal energy flux related only to the thermal conductivity (lower dashed: $\eta$=40\%; dotted line: $\eta$=15\%).}
\end{center}
\end{figure*}

\begin{figure*}[!ht]
\begin{center}
\includegraphics[width=0.49\columnwidth]{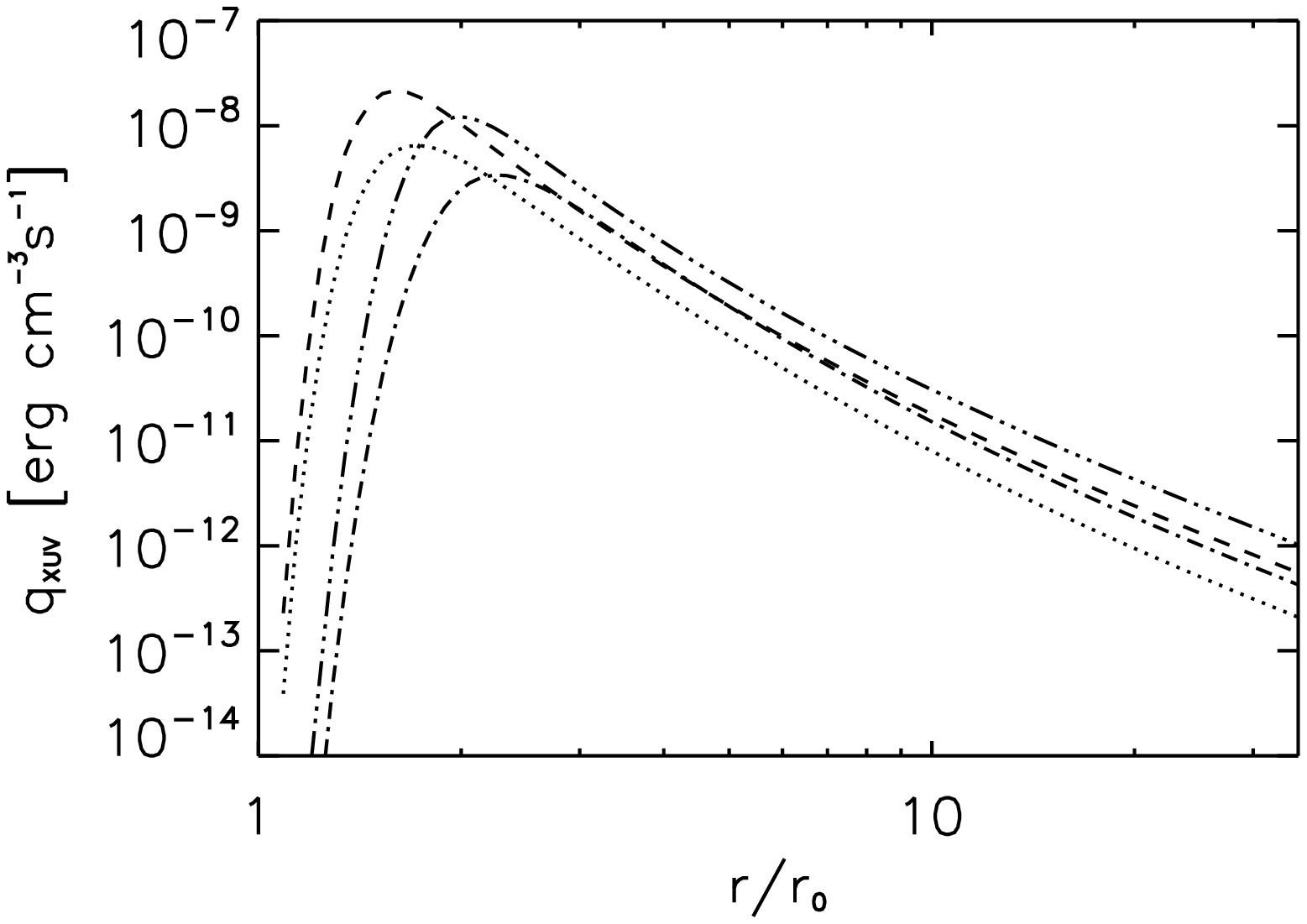}
\includegraphics[width=0.49\columnwidth]{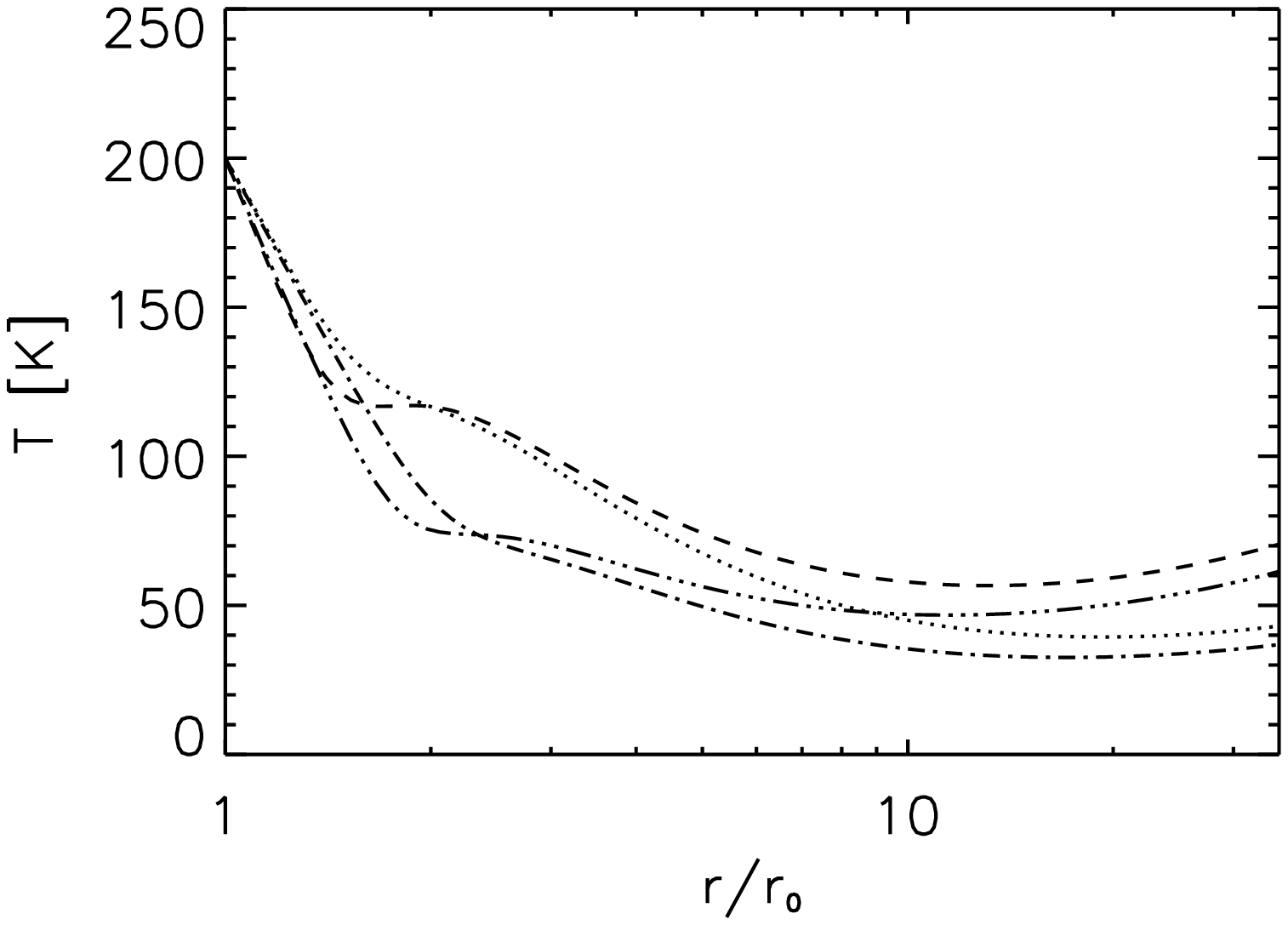}
\includegraphics[width=0.49\columnwidth]{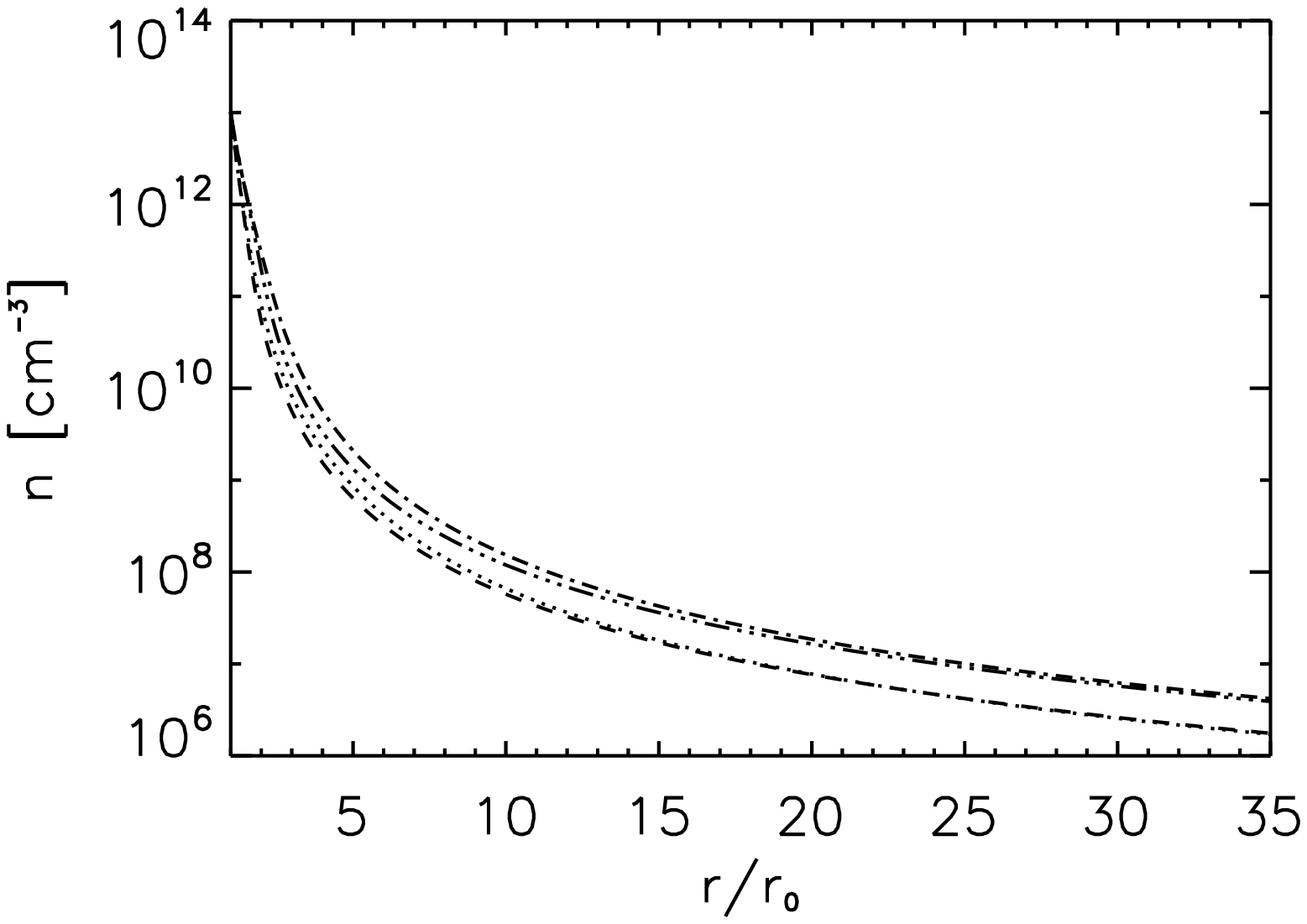}
\includegraphics[width=0.49\columnwidth]{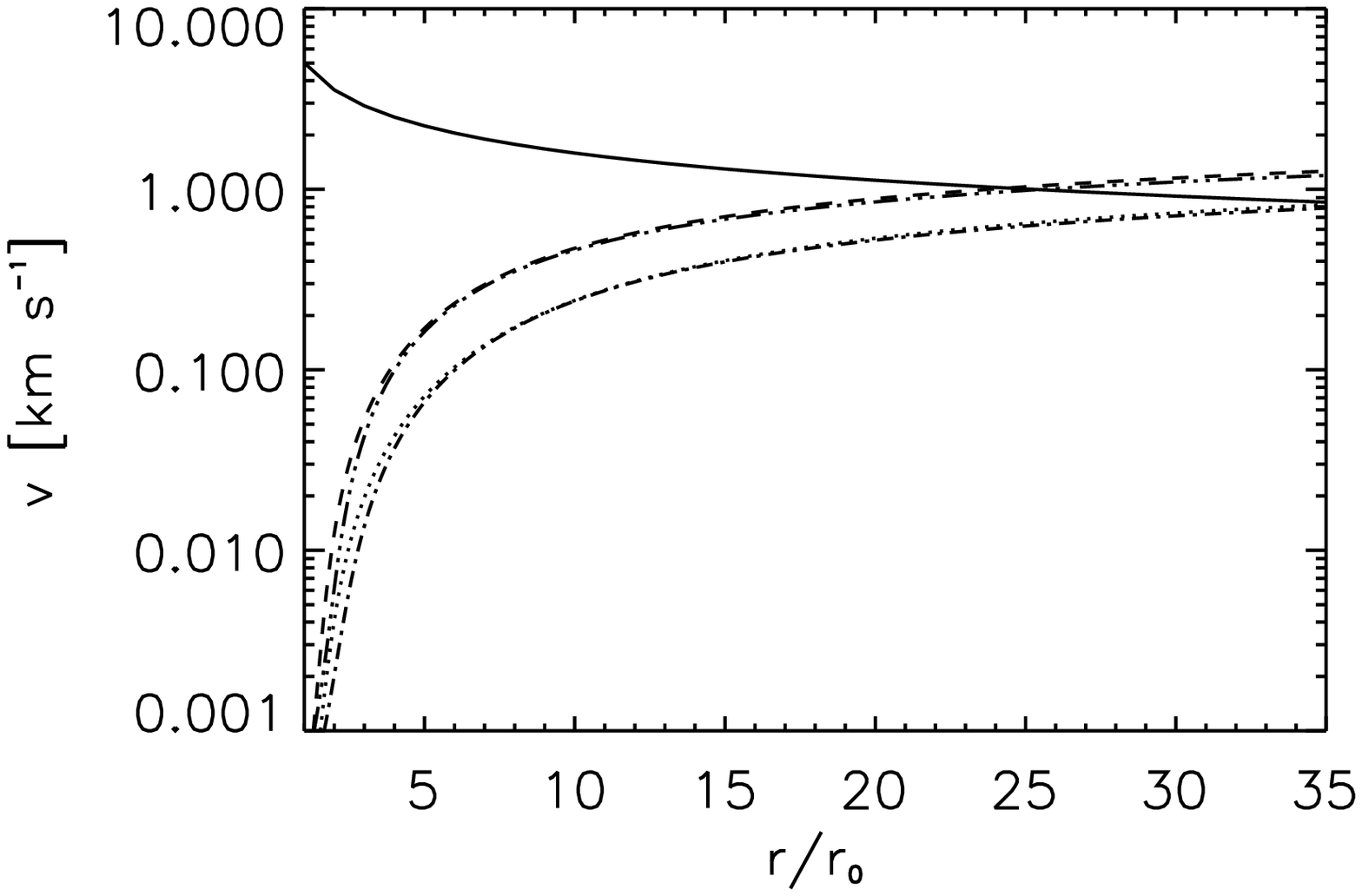}
\caption{Example of the XUV volume heating production rate (top left), temperature profile (top right), density profile (bottom left) and the velocity profile (bottom right)
for a hydrogen-rich martian upper atmosphere with $z_0$=100 km by assuming a heating efficiency of 15\% (dotted lines) and 40 \% (dashed lines)
and a temperature $T_0$ at the base of the thermosphere of 200 K
as a function of distance in planetary radii for a hydrogen-dominated upper atmosphere at Mars, that is exposed to a 100 time higher XUV flux compared to today's solar value.
The solid line shown in the velocity profiles corresponds to the escape velocity $v_{\rm esc}$ as a function of distance.
The dashed-dotted lines ($\eta=15$\%) and the dashed-dotted-dotted-dotted
lines ($\eta=40$\%) correspond to similar profiles but with $z_0=1000$ km.
The hydrogen atoms reach the escape velocity below the theoretical exobase level at a location of $\sim$35 $r_{\rm 0}$ for $\eta$=15\% and
at $\sim$24 $r_{\rm pl}$ for $\eta$=40\%.}
\end{center}
\end{figure*}
\begin{table*}
\caption{Modeled atmospheric parameters and thermal hydrogen atom escape rates $L_{\rm th}$ corresponding to a 100 times higher XUV flux compared to today's Sun
at the critical distance $r_{\rm c}\leq r_{\rm exo}$, where the dynamically outward flowing hydrogen dominated bulk atmosphere reaches (sonic speed)
above the planetary surface and two heating efficiencies $\eta$ of 15 \% and 40\%.}
\begin{center}
\begin{tabular}{l|ccccccc}
CASES & $\eta$ [\%]   & $z_0$ [km] & $r_{\rm XUV_{\rm eff}}$ [$r_0$]  & $r_{\rm c}$ [$r_0$]  & $n_{\rm c}$ [cm$^{-3}$]  & $T_{\rm c}$ [K] &  $L_{\rm th}$ [s$^{-1}$] \\\hline
CI   & 15             &100         &  3.4                         &   32.5         &  2$\times 10^{6}$       & 40        &  $1.8\times10^{32}$  \\
CII  & 15             &1000        &  4.5                    &   30           &  6.2$\times 10^{6}$     & 36        &  $7.0\times10^{32}$ \\
CIII & 40             &100         &  3.2                   &   21           &  6.5$\times 10^{6}$     & 60        &  $3.0\times10^{32}$ \\
CIV  & 40             &1000        &  4.2                  &   20           &  1.7$\times 10^{7}$     & 50        &  $1.0\times10^{33}$ \\\hline
\end{tabular}
\end{center}
\end{table*}
Fig. 6 shows examples of the XUV volume heating rate and the corresponding upper atmosphere structure of a
hydrogen dominated upper atmosphere of early Mars for a heating efficiency $\eta$ of 15 and 40 \% with $T_0=200$ K,
and $n_0=10^{13}$ cm$^{-3}$, which is exposed to a XUV flux which is
100 times higher at the planet's orbit compared to that of the present Sun and assumed mesopause locations at
100 km and 1000 km.
Under these assumptions the bulk atmosphere reaches the escape velocity $v_{\rm esc}$ at about 35$r_0$ and 24$r_0$
for heating efficiencies $\eta$ of 15\% and 40\%, respectively.
One can see from the volume heating rate $q_{\rm XUV}$ and the connected temperature profile
that the XUV deposition peak occurs above 1.5$r_0$ for $z_0=100$ km and at $\sim 2R_0$ if $z_0=1000$ km.
This can also be seen in the temperature
profiles, which decrease first due to adiabatic cooling until the high XUV flux of the young Sun balances the
cooling process due the to XUV heating, resulting in the more or less constant temperature profile between $\sim$5--35$r_{\rm pl}$ of $\sim$50--70 K.
One can also see that for a heating efficiency $\eta$ of 40\% the adiabatic cooling is stronger at distances that are $\leq 2.0r_0$.
The corresponding temperature drop is also larger for an $\eta$ of 40\% compared to that of 15\%.
For larger distances $r>2r_0$, the energy absorption is larger and in the case of 40\% efficiency, the additional heating
exceeds the cooling. Therefore, the temperature decrease is less pronounced for large distances in the case of higher heating efficiencies
compared to the lower value of $\eta$=15\%.

Table 3 shows the thermal hydrogen atom escape rates and relevant atmospheric
parameters at the critical distance
where the bulk atmosphere reaches sonic speed for a lower and higher heating efficiency $\eta$ of
15\% and 40\% and for $z_0$ at 100 and 1000 km altitude. The temperature $T_0$ and the
number density $n_0$ is assumed to be 200 K and $10^{13}$ cm$^{-3}$ in all four cases.
One can see from Table 3 that depending on $z_0$ the thermal hydrogen escape rates can reach values between $\sim 2\times 10^{32}$
and $\sim 10^{33}$ H atoms per second. The present time thermal hydrogen atom escape from Mars
by the classical Jeans escape is about $\sim1.5\times 10^{26}$ s$^{-1}$ (e.g., Lammer et al., 2008), which indicates that the
thermal escape of hydrogen from Mars' protoatmosphere could have been up to $\sim$6--7 orders of magnitude higher.
\subsection{Escape of the nebula captured hydrogen envelope}
By knowing the escape rate of hydrogen atoms we can now estimate the loss of the expected nebula-based hydrogen
envelope from proto-Mars. If we use the most massive captured hydrogen envelope shown in Table 1 of $\sim 6.5 \times 10^{22}$ g, corresponding to a luminosity of
$\sim2.4 \times 10^{22}$ erg s$^{-1}$ and a dust grain depletion factor $f$ of 0.1, the envelope would be lost during $\sim 1.3$--$7.5$ Myr. The escape time span depends
on the heating efficiency and the location distance of the lower thermosphere. A more realistic captured atmosphere with a mass of $\sim 5\times 10^{21}$ g would be
lost in $\sim 0.1$--$0.5$ Myr. From these escape estimates one can conclude that a captured nebular-based hydrogen envelope should have been lost very fast from the planet
after the nebula dissipated. If the radius $r_0$ in the nebula captured hydrogen envelope was at further distances compared to our assumed values,
then the escape rates would be higher.
\subsection{Escape of the catastrophically outgassed steam atmosphere}
According to the outgassing of the magma ocean depth dependent steam atmospheres shown in Table 2, even the deepest and most volatile-rich case
completes solidification and degassing in $\leq2\times 10^5$ years. The heat loss from the small planetary body is fast enough to
allow rapid solidification in a convecting magma ocean. Theoretical studies by
Elkins-Tanton (2008) showed that one can expect that the volatiles are likely to be released
toward the end of solidification of the magma ocean in a ``burst''.

The applied magma ocean model of Elkins-Tanton (2008) and the
related results discussed in Sect. 2.2 predict a surface temperature of $\geq$ 800 K at
the end of solidification, which lies above the condensation temperature for H$_2$O of $\sim$645 K.
If there is a solid-state mantle overturn,
there will be a big temperature jump after $\sim$2--4 Myr, when the hot mantle cumulates rise up in Mars because of their buoyancy, and advect their
great heat with them. According to Brasser (2013) Mars' most likely finished its accretion or remained as
a planetary embryo when the surrounding nebula was still present around the martian orbit location in its later stages. If this was the case the catastrophically outgassed
volatiles could easily build up rapidly around the rocky embryo.
As soon as this catastrophically outgassed steam atmosphere
was released from the nebula the efficient escape of the atmosphere which was driven by the high XUV flux of the young Sun began.

According to Lebrun et al. (2013), who studied the thermal evolution of an early
martian magma ocean in interaction with a catastrophically outgassed $\sim 43$ bar
H$_2$O and $\sim 14$ bar CO$_2$ steam atmosphere, water vapor would start to condense into liquid H$_2$O after $\sim$0.1 Myr.
On the other hand, such a fast cooling of the steam atmosphere contradicts the isotopic analysis of martian SNC meteorites by Debaille et al. (2007),
where analyzed data can be best explained by a progressive crystallization of a magma ocean with a duration of up to $\sim$ 100 Myr.
Therefore, Lebrun et al. (2013) suggest that frequent impacts of large
planetesimals and small embryos, which have been not included in their study, could have kept the surface during longer times warmer.
This suggestion is quite logical because one can also expect that during the first 100 Myr after the origin of the Solar System the young
planets have been frequently hit by large impactors (e.g., Abe and Matsui, 1985; 1988; Genda and Abe, 2005; Lammer et al., 2013a),
which may have enhanced the input energy flux above the value which is defined by the solar flux alone. In such a case one will obtain
a hotter surface that prevent atmospheric H$_2$O vapor from condensing (e.g., Hayashi et al., 1979; Genda and Abe, 2005; Lammer et al., 2012; Lammer, 2013;
Lebrun et al., 2013).

One should also note that for the surface temperatures of $\sim$500 K, which are expected during the ``Mush'' stage (Lebrun et al., 2013),
according to Kasting (1988) one can also expect water vapor mixing ratios at the mesopause level near to 1. For that reason H$_2$O will continue
to escape effectively, even if there are periods of liquid water on the planet's surface. However, the mesopause level will be closer
to the planet's surface and the escape rates will be reduced and may have values which correspond to case CI in Table 3.

In the outgassed steam atmosphere, the H$_2$O molecules in the upper atmosphere will be dissociated by the high XUV flux of the young Sun
and by frequently occurring impacts in the lower thermosphere (e.g., Chassefi\`{e}re, 1996b; Lammer et al., 2012; Lammer, 2013). Tian et al. (2009)
showed that for XUV fluxes which
are $>$ 10 times that of today's Sun, CO$_2$ and/or CH$_4$ molecules in the martian upper atmosphere will also be destroyed, so that C atoms can escape
similar to O atoms with escape flux values which are $\geq10^{11}$ cm$^{-2}$ s$^{-1}$. From this study one can expect that for an
XUV flux which is $\sim$100 times stronger than the present solar value most CO$_2$ and/or CH$_4$ molecules will be dissociated
as soon as they are exposed to the high XUV radiation. Therefore, one can assume that O and C atoms should also populate the lower hydrogen dominated
thermosphere so that they can be dragged by the dynamically outward flowing hydrogen atom flux (Zahnle and Kasting, 1986; Chassefi\`{e}re, 1996a; 1996b;
Hunten et al., 1987; Lammer et al., 2012; 2013a).
\begin{figure*}[!ht]
\begin{center}
\includegraphics[width=0.49\columnwidth]{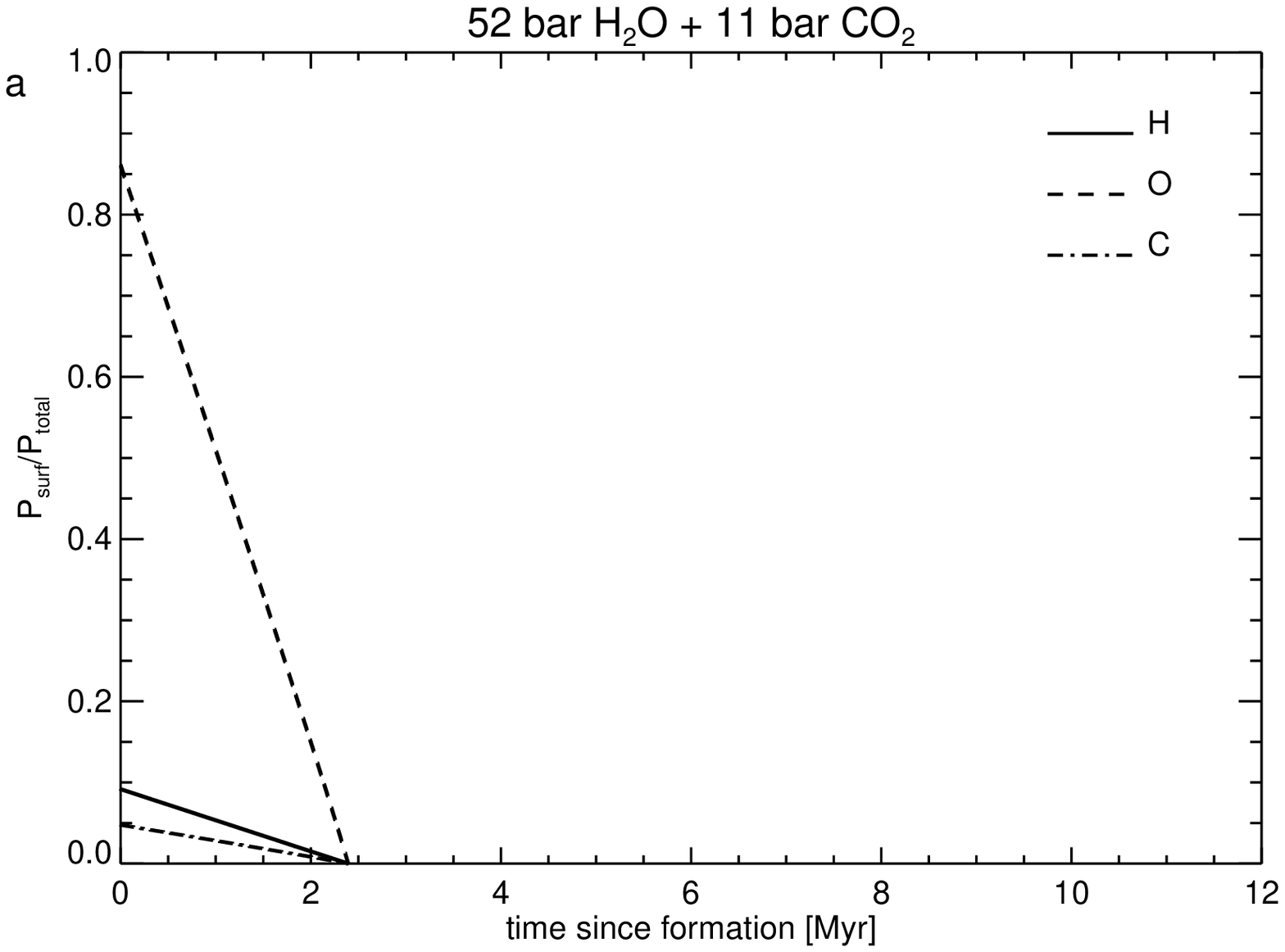}
\includegraphics[width=0.49\columnwidth]{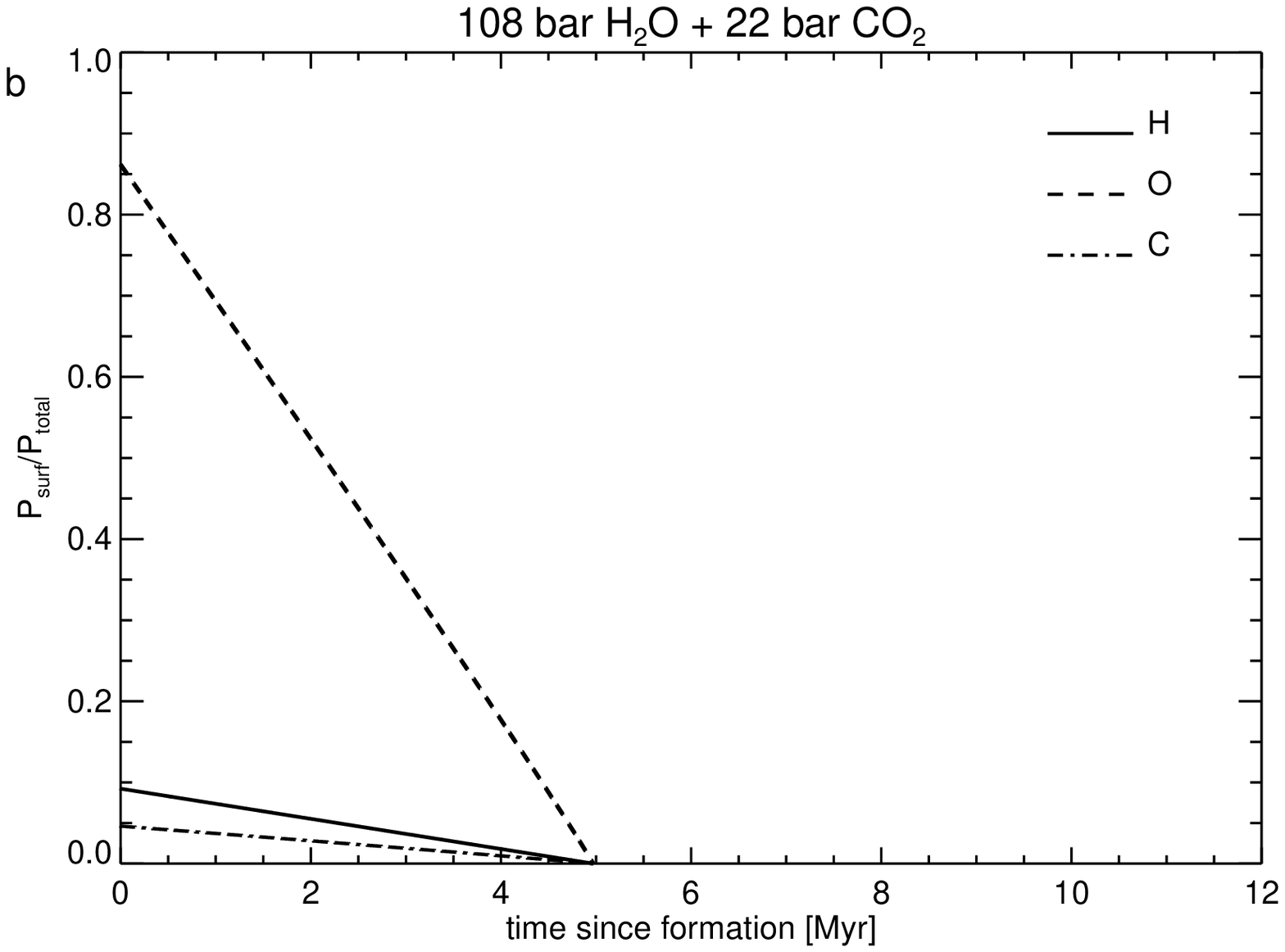}
\includegraphics[width=0.49\columnwidth]{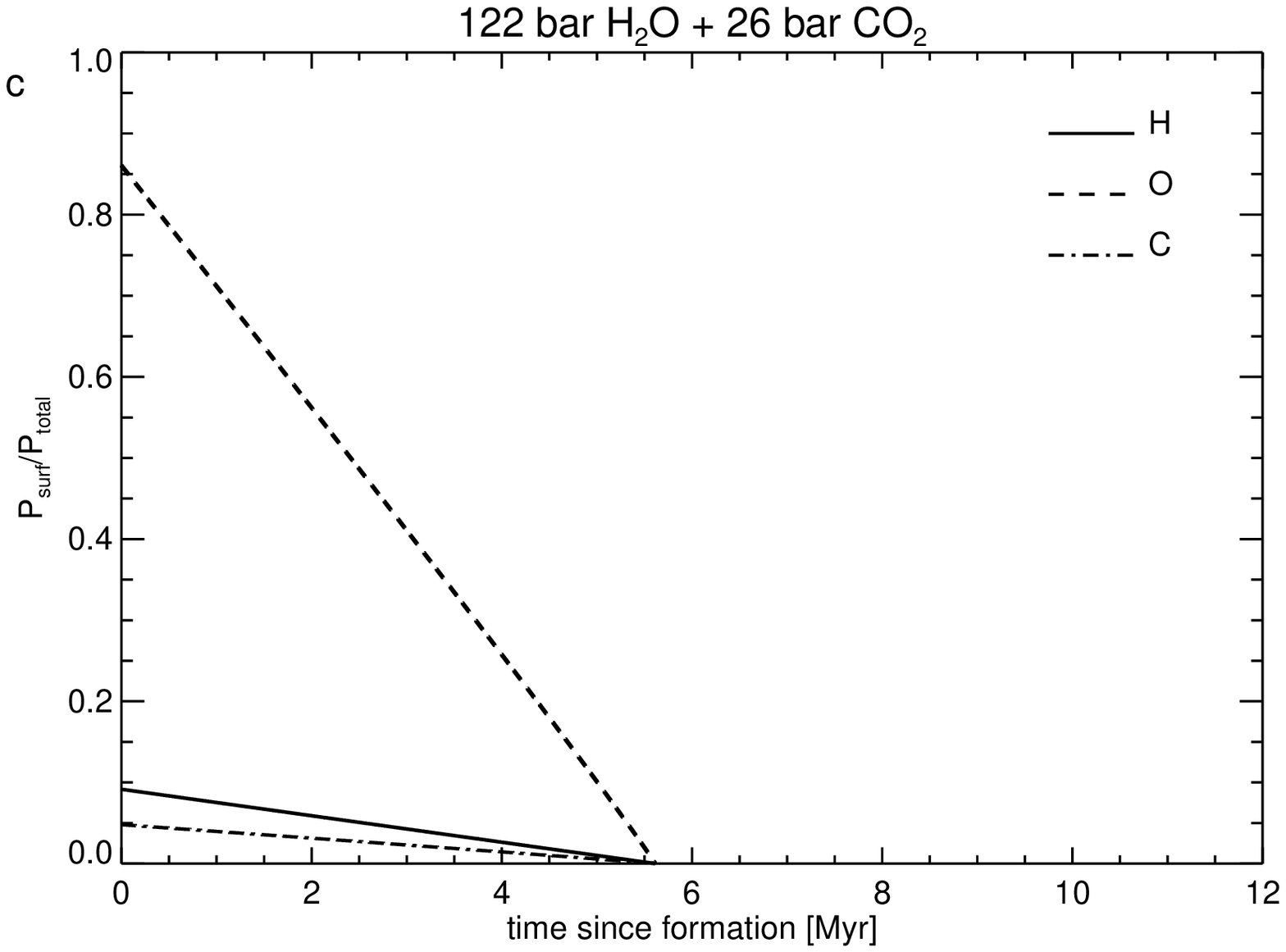}
\includegraphics[width=0.49\columnwidth]{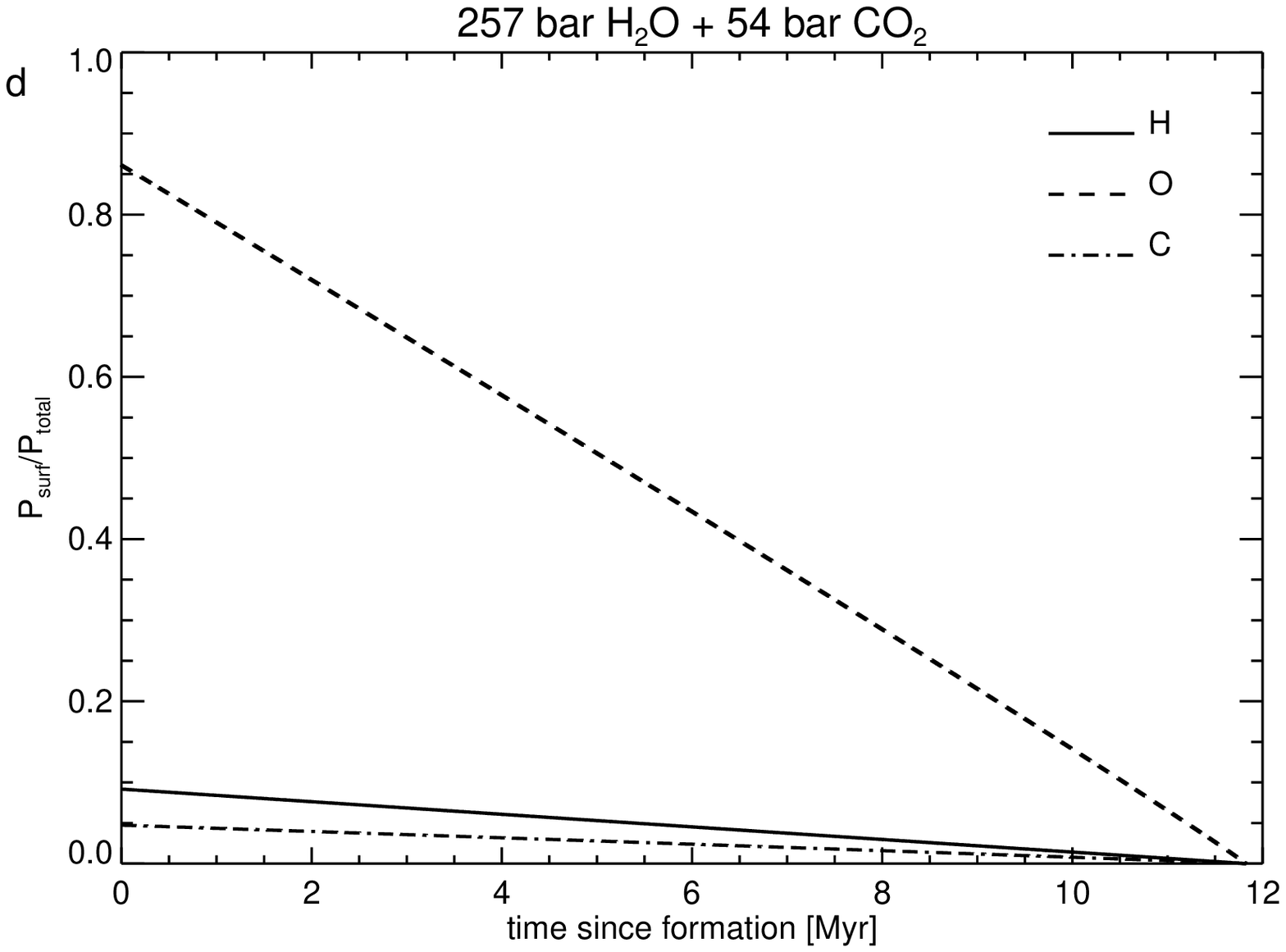}
\caption{Temporal evolution of the partial surface pressures $P_\mathrm{surf}$ of H, O, and C normalized to the total initial surface pressure $P_\mathrm{total}$ for the four
compositions of outgassed atmospheres described in Table 2. The hydrogen inventory evolves assuming a constant escape rate and parameters according to CI in Table 3
valid for 100 XUV. Both O and C are dragged along with the escaping H.}
\end{center}
\end{figure*}

\begin{figure*}[!ht]
\begin{center}
\includegraphics[width=0.49\columnwidth]{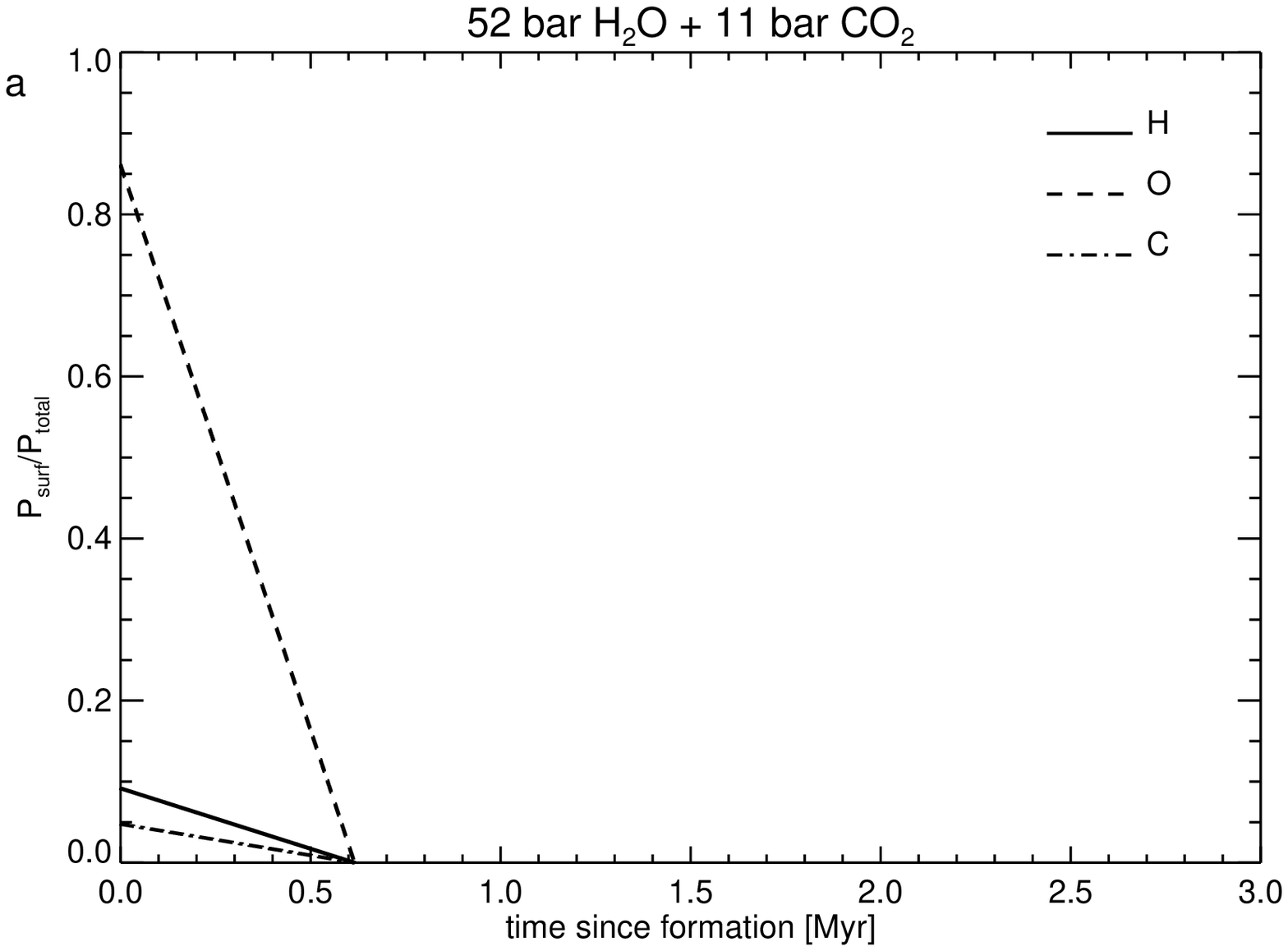}
\includegraphics[width=0.49\columnwidth]{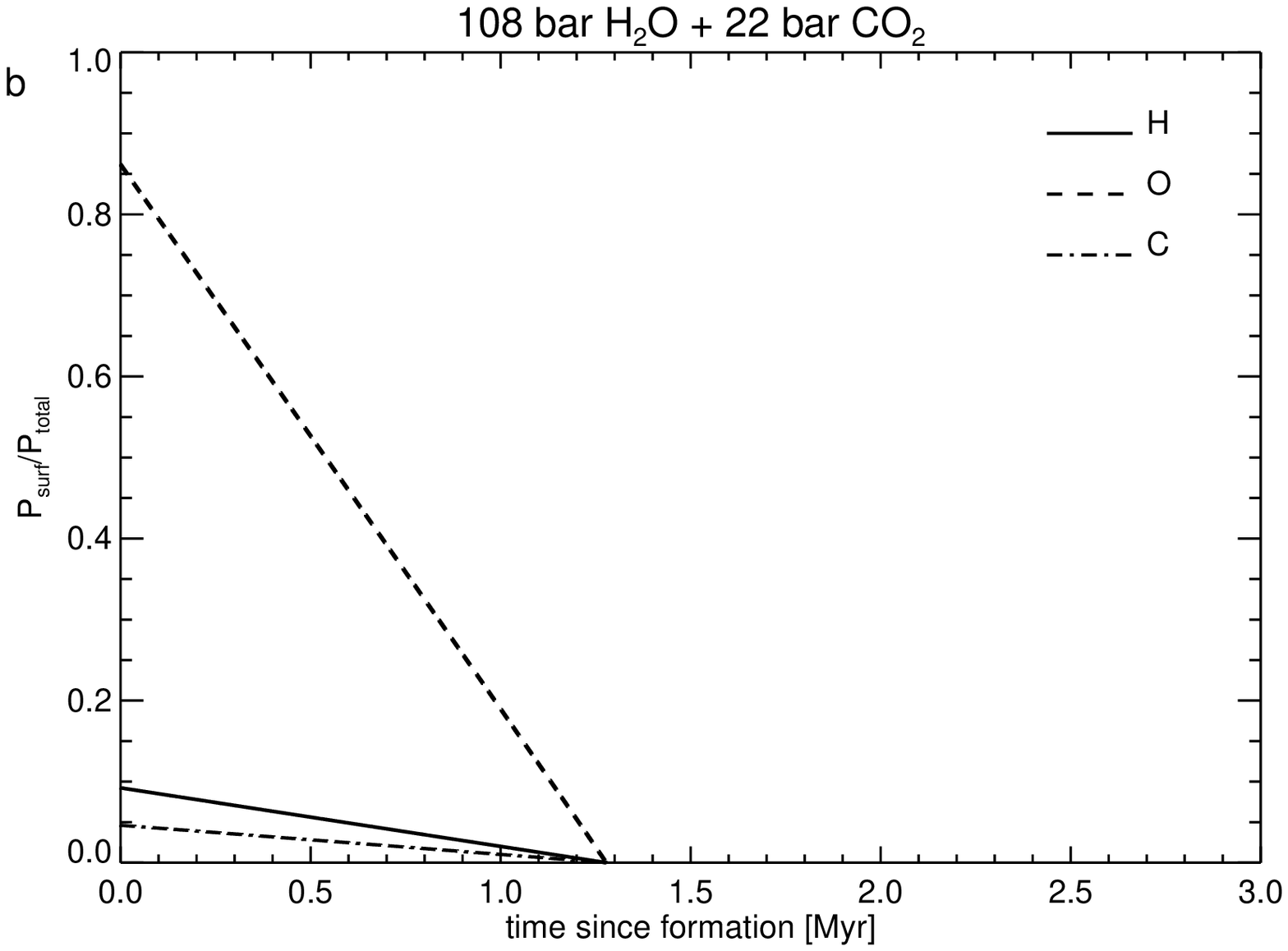}
\includegraphics[width=0.49\columnwidth]{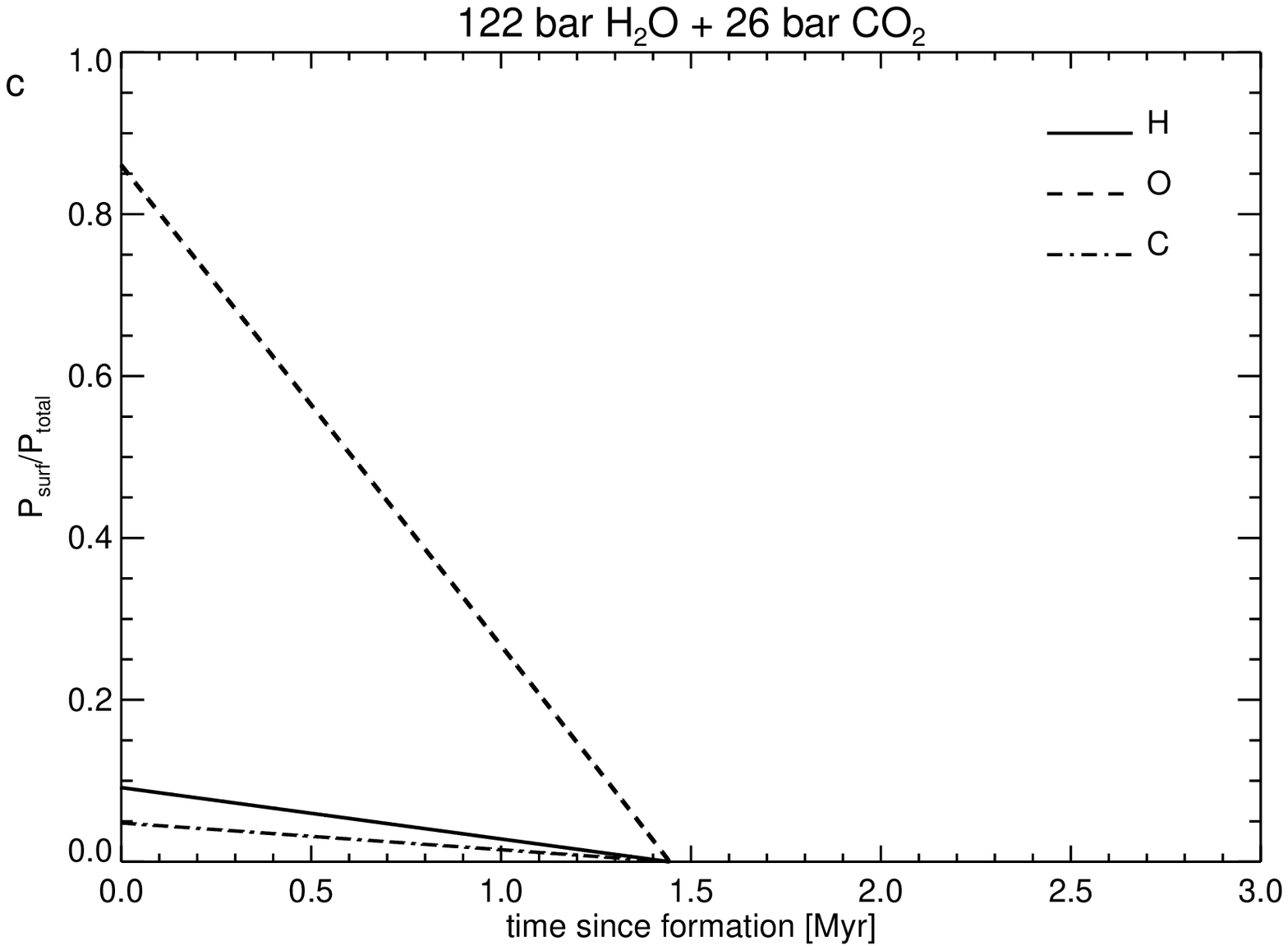}
\includegraphics[width=0.49\columnwidth]{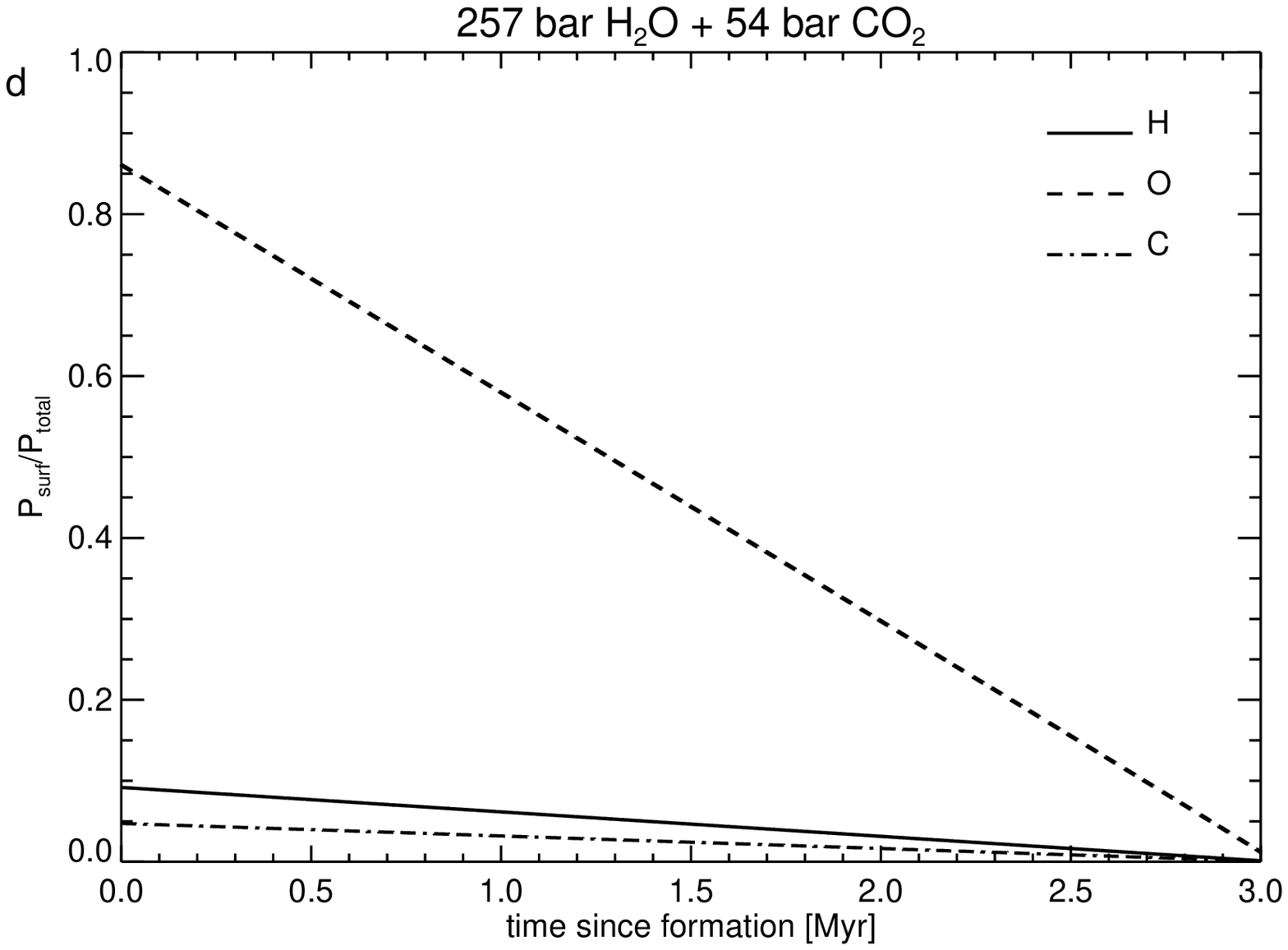}
\caption{Temporal evolution of the partial surface pressures $P_\mathrm{surf}$ of H, O, and C normalized to the total initial surface pressure $P_\mathrm{total}$ for the four
compositions of outgassed atmospheres described in Table 2. The hydrogen inventory evolves assuming a constant escape rate and parameters according to CII in Table 3
valid for 100 XUV. Both O and C are dragged along with the escaping H.}
\end{center}
\end{figure*}

\begin{figure*}[!ht]
\begin{center}
\includegraphics[width=0.49\columnwidth]{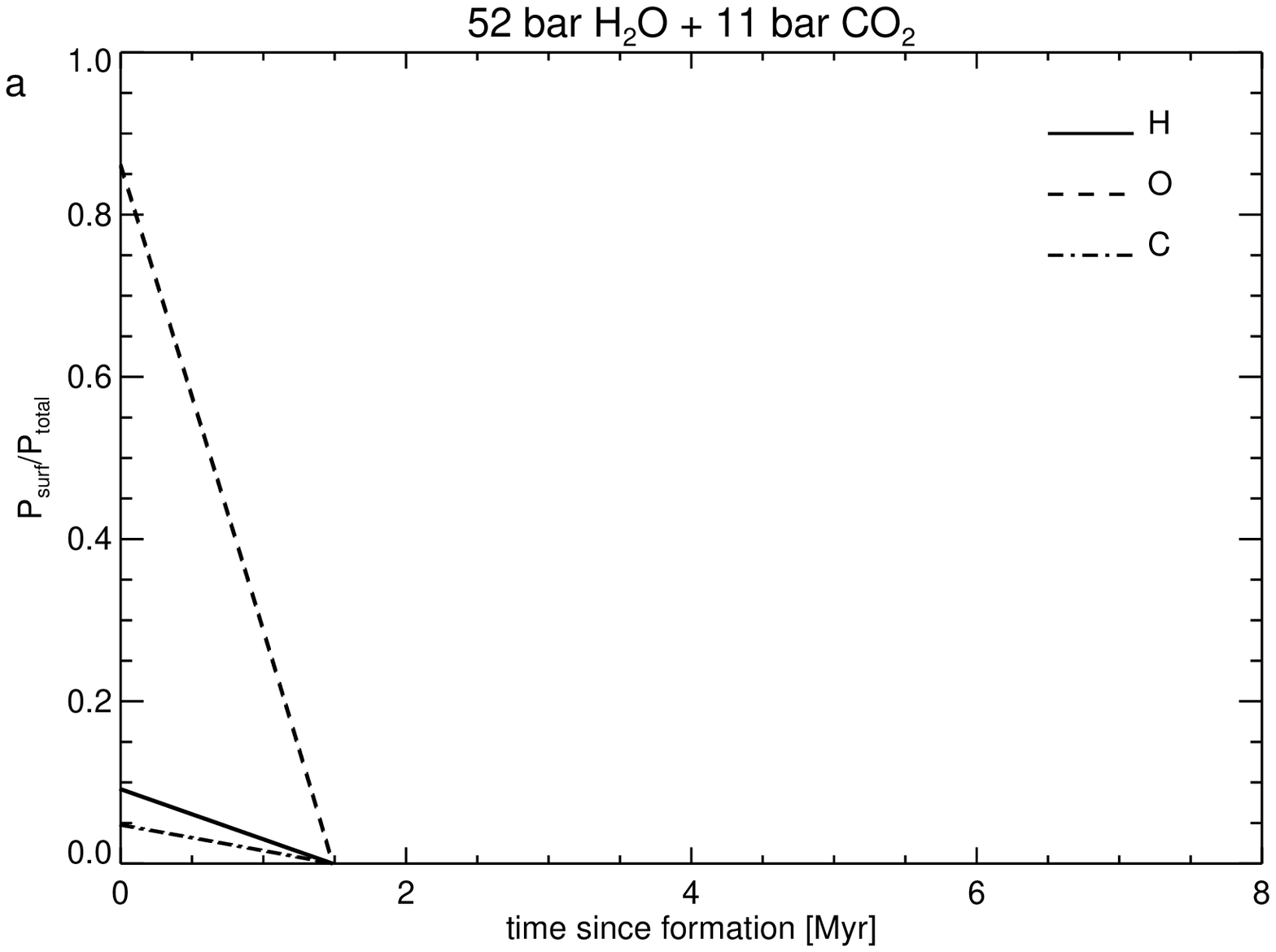}
\includegraphics[width=0.49\columnwidth]{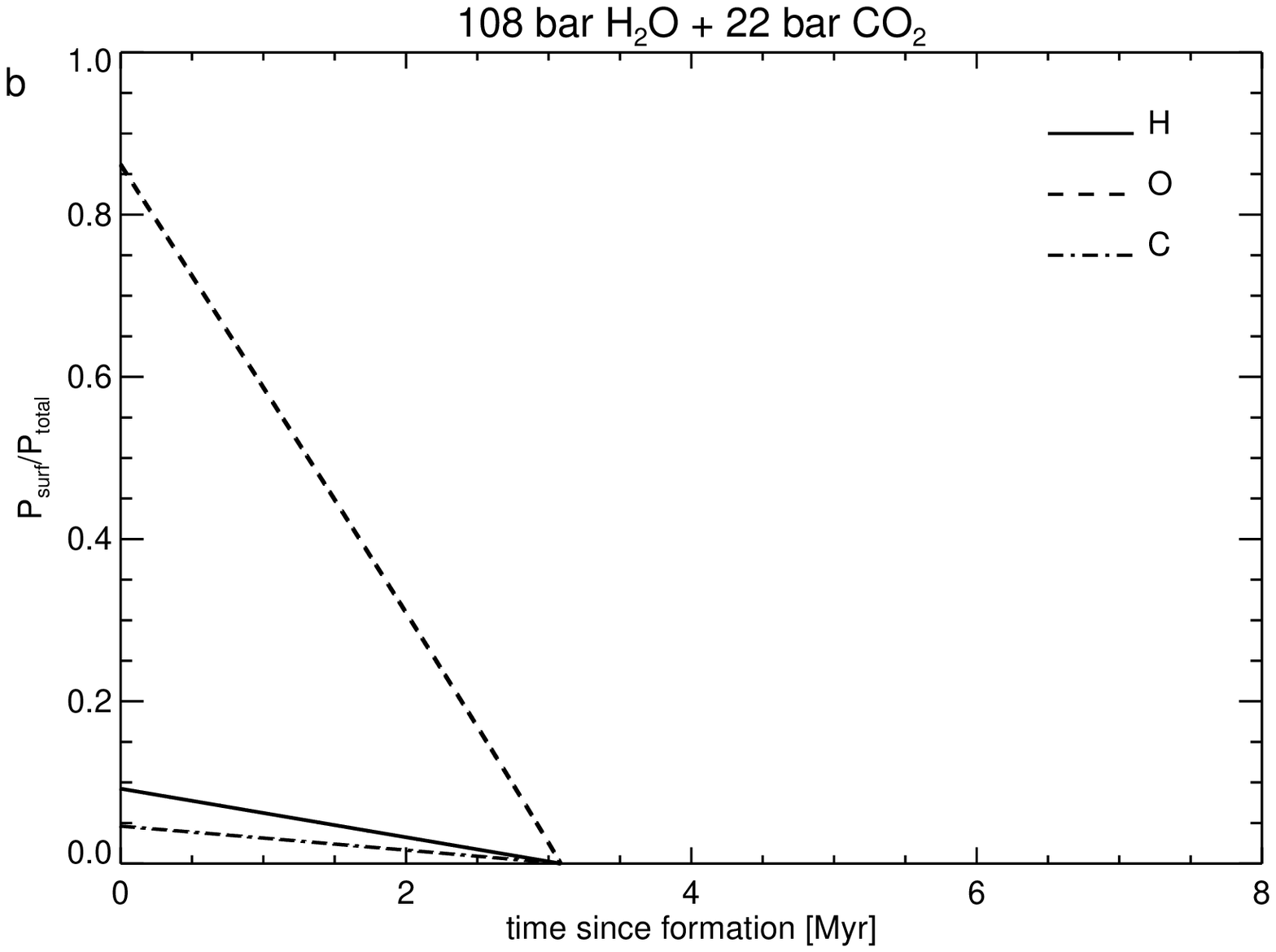}
\includegraphics[width=0.49\columnwidth]{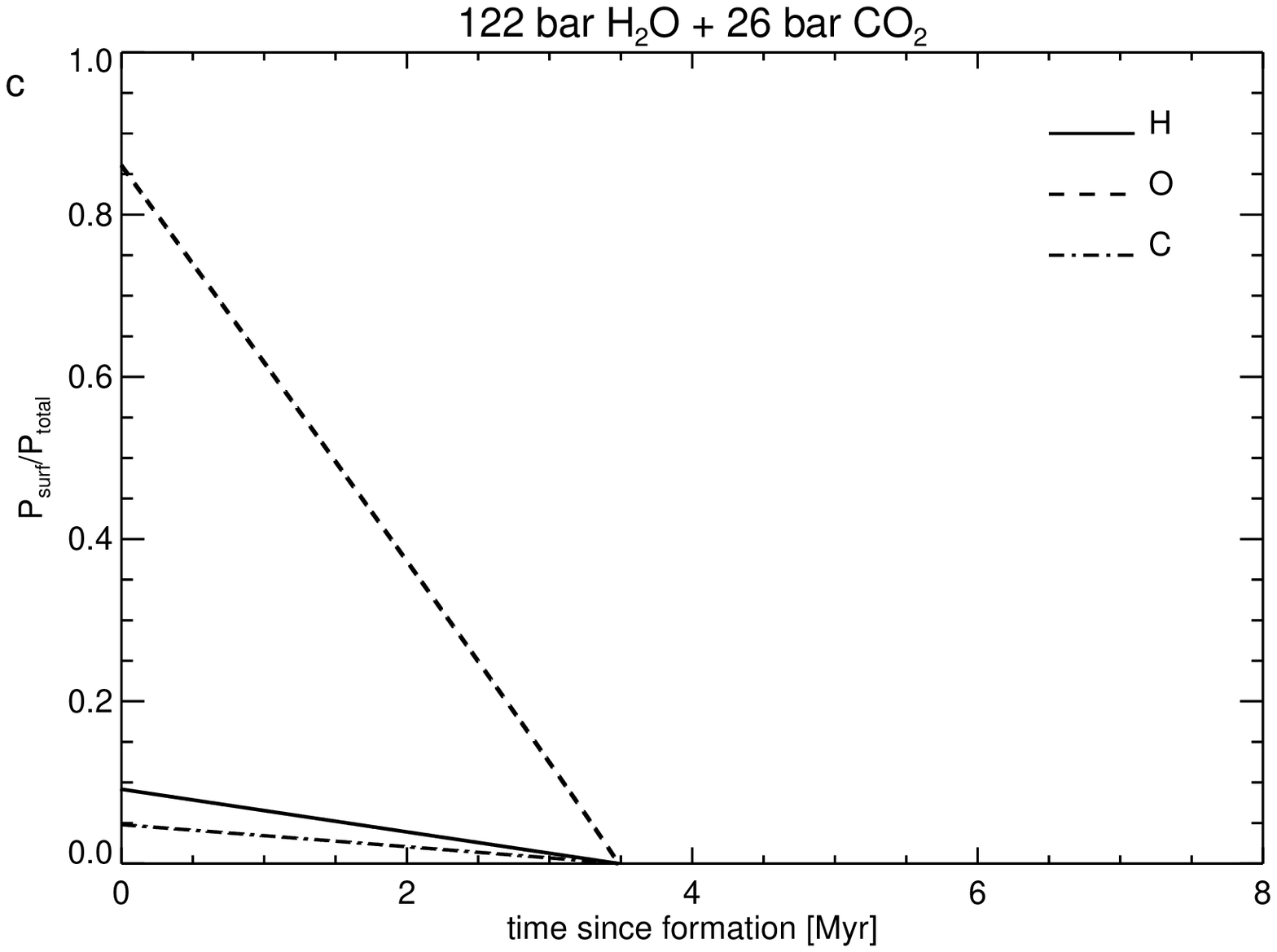}
\includegraphics[width=0.49\columnwidth]{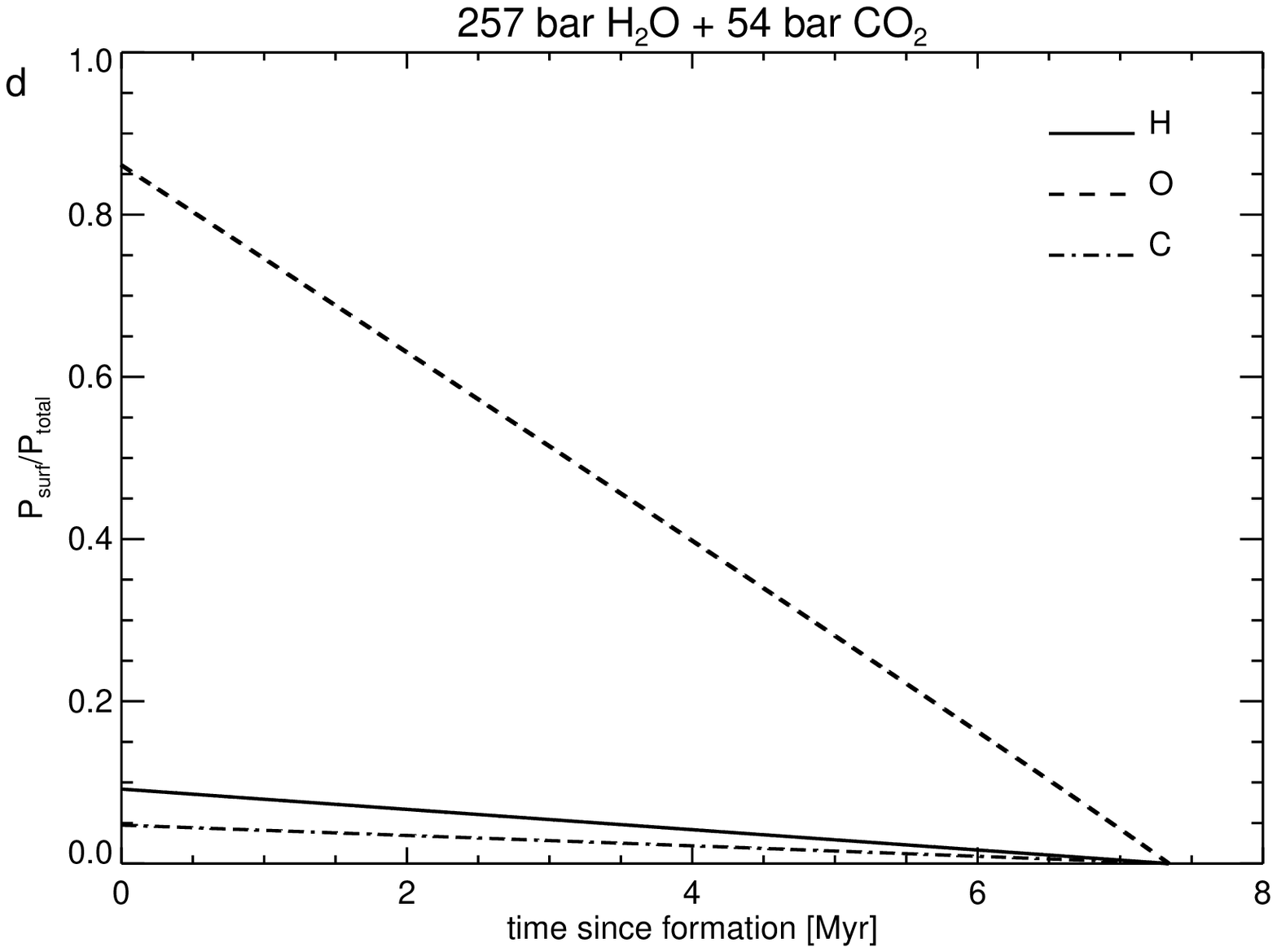}
\caption{Temporal evolution of the partial surface pressures $P_\mathrm{surf}$ of H, O, and C normalized to the total initial surface pressure $P_\mathrm{total}$ for the four
compositions of outgassed atmospheres described in Table 2. The hydrogen inventory evolves assuming a constant escape rate and parameters according to CIII in Table 3
valid for 100 XUV. Both O and C are dragged along with the escaping H.}
\end{center}
\end{figure*}

\begin{figure*}[!ht]
\begin{center}
\includegraphics[width=0.49\columnwidth]{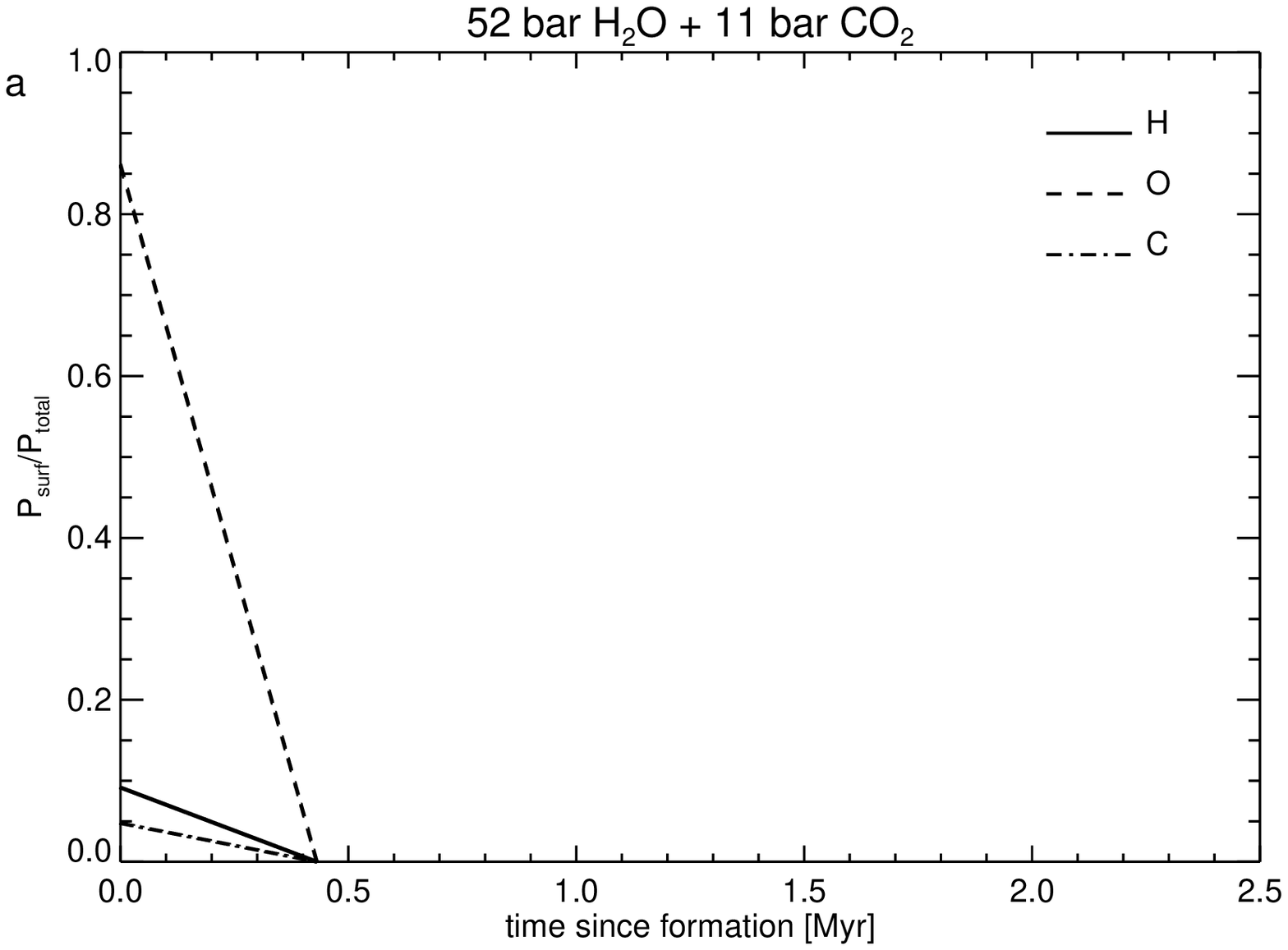}
\includegraphics[width=0.49\columnwidth]{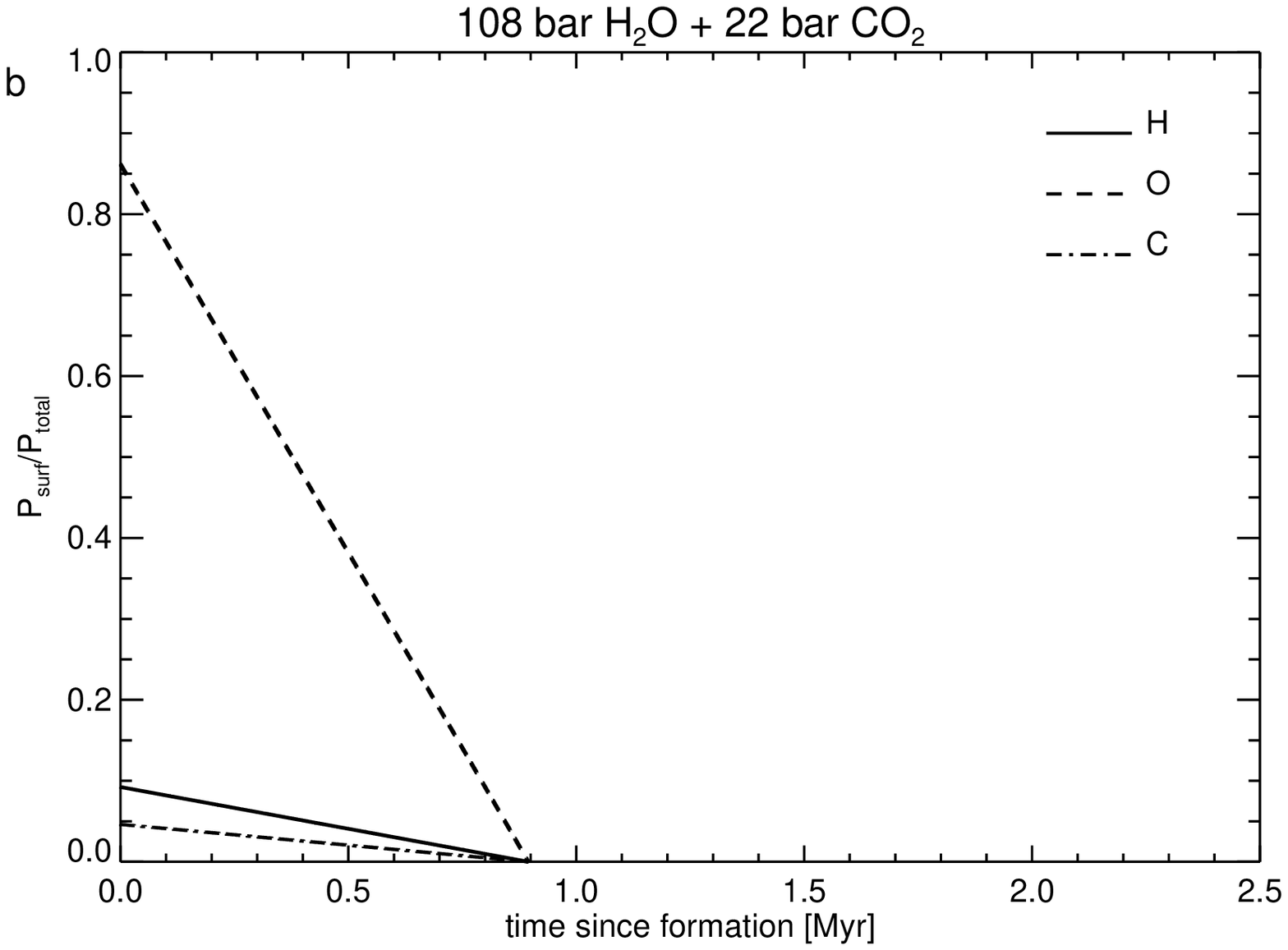}
\includegraphics[width=0.49\columnwidth]{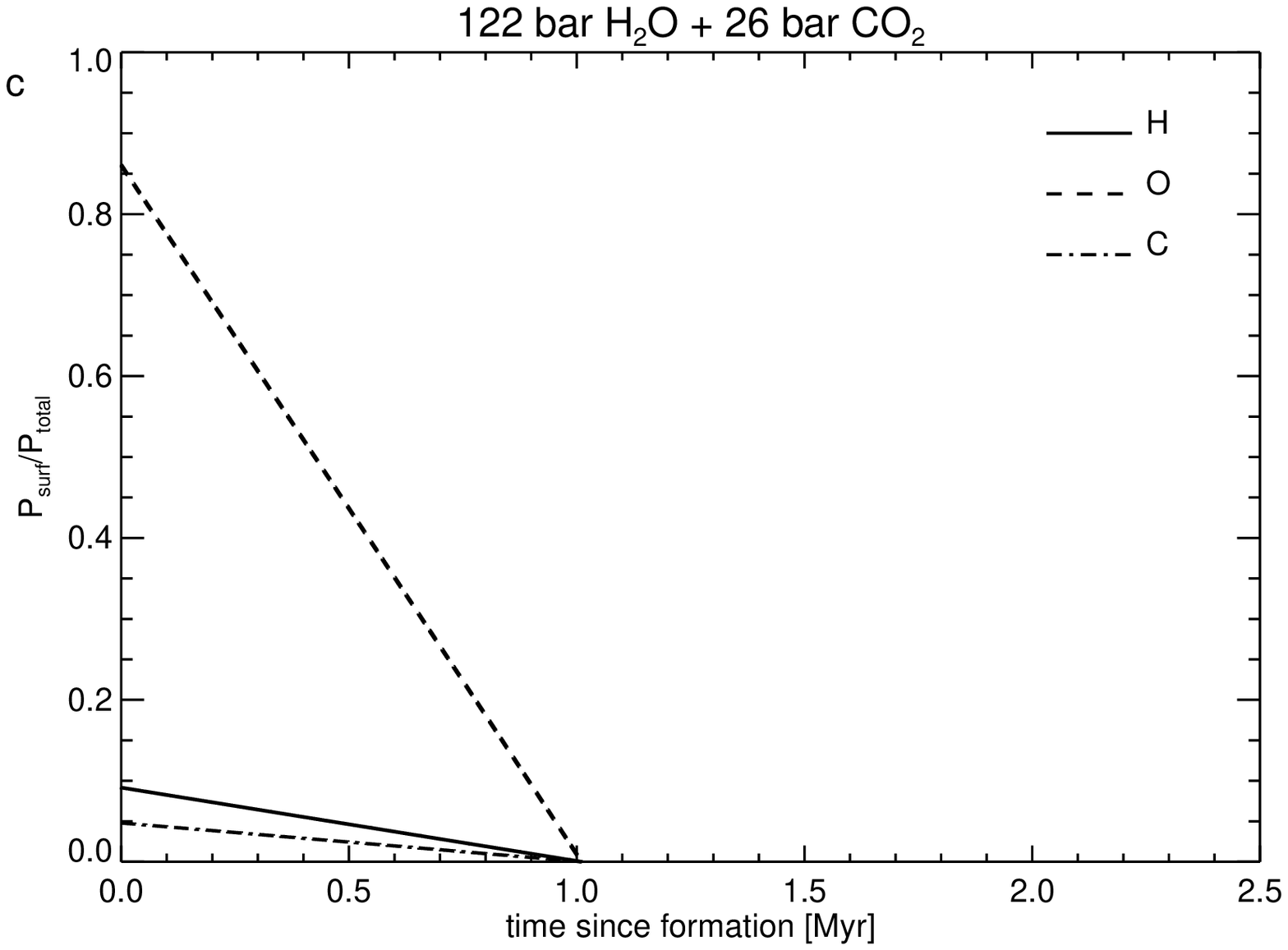}
\includegraphics[width=0.49\columnwidth]{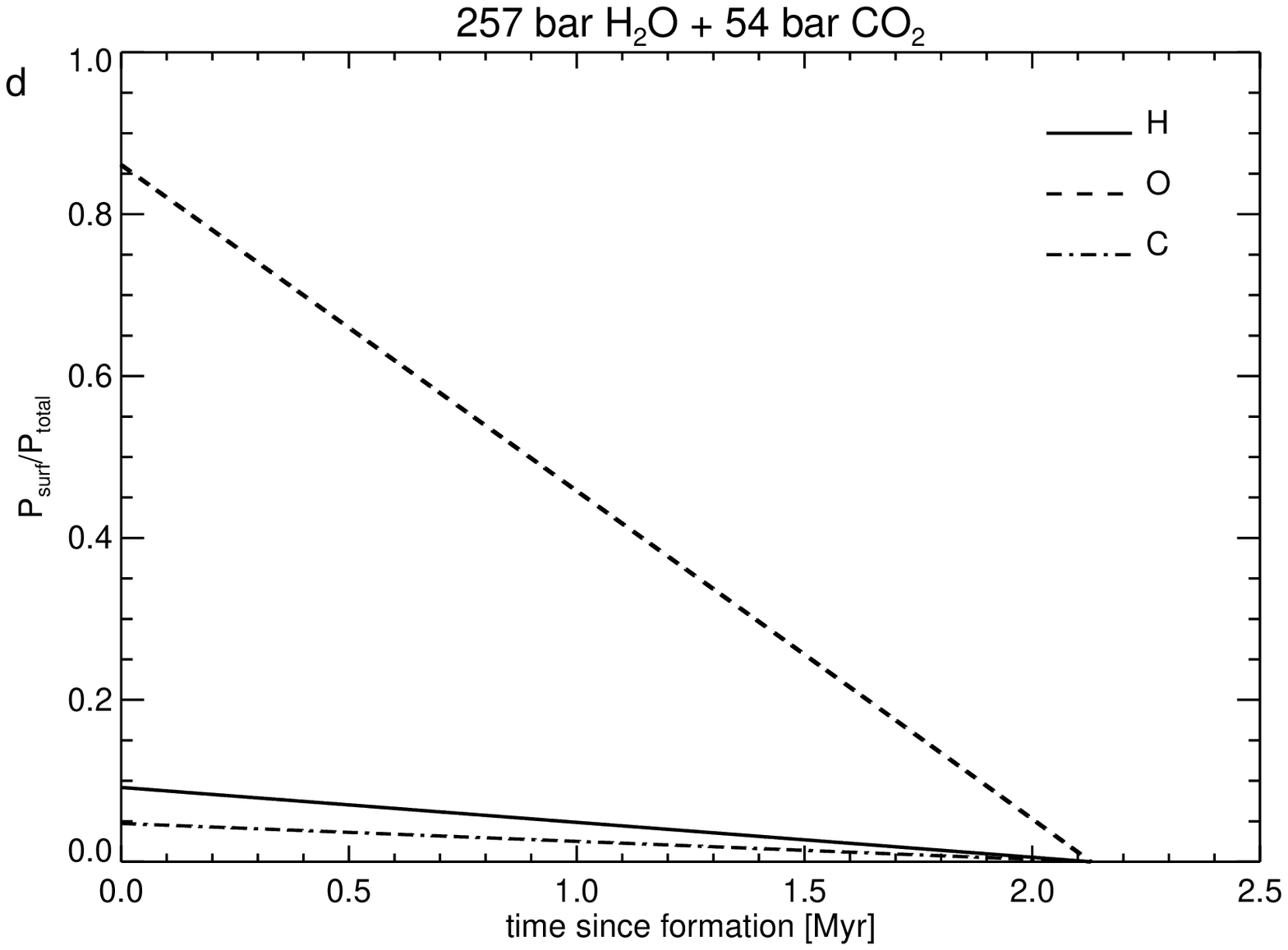}
\caption{Temporal evolution of the partial surface pressures $P_\mathrm{surf}$ of H, O, and C normalized to the total initial surface pressure $P_\mathrm{total}$ for the four
compositions of outgassed atmospheres described in Table 2. The hydrogen inventory evolves assuming a constant escape rate and parameters according to CIV in Table 3
valid for 100 XUV. Both O and C are dragged along with the escaping H.}
\end{center}
\end{figure*}
As initial amount and composition of the outgassed atmosphere we adopt the four cases presented in Table 2.
With the given partial surface pressures of H$_2$O and CO$_2$ and assuming that all molecules are dissociated
under the high XUV flux of the young Sun, we calculate the initial inventories of atomic H, O, and C. For all
four cases atomic hydrogen is the most abundant species ($N_\mathrm{H}/N=0.61$), followed by oxygen ($N_\mathrm{O}/N=0.36$),
whereas C is just a minor constituent ($N_\mathrm{C}/N=0.03$). Hydrogen is assumed to escape at rates given in Table 3.
The fractionation factors $x_{\rm i}=L_{\rm i}/(L_\mathrm{H}f_{\rm i})$ for an escaping atmosphere composed of two major
(here H, O) and several minor species (here only one, namely C) are given by Eqs. 35 and 36 of Zahnle and Kasting (1986),
where $f_{\rm i}=n_{\rm i}/n_\mathrm{H}=N_{\rm i}/N_\mathrm{H}$ is the mixing ratio with respect to H and $L_{\rm i}$ are
the escape fluxes of the heavy species $i$ given in s$^{-1}$. Using the definition of $x$ the escape fluxes of O and C can
then by written as
\begin{equation}\label{eq:LO}
	L_\mathrm{O} = L_\mathrm{H} f_\mathrm{O} x_\mathrm{O} = L_\mathrm{H} f_\mathrm{O}
	\left(1-\frac{\mu_\mathrm{O}-1}
	{\mu_\mathrm{O}\Phi_\mathrm{O}} \frac{1}{1+f_\mathrm{O}}\right)
\end{equation}
\begin{equation}\label{eq:LC}
	L_\mathrm{C} = L_\mathrm{H} f_\mathrm{C} \frac{1-\frac{\mu_\mathrm{C}-1}
	 {\mu_\mathrm{C}\Phi_\mathrm{C}}+\frac{b_\mathrm{HC}}{b_\mathrm{OC}}f_\mathrm{O}x_\mathrm{O}+
	\frac{b_\mathrm{HC}}{b_\mathrm{HO}}\frac{f_\mathrm{O}(1+f_\mathrm{O})
	 (1-x_\mathrm{O})}{\mu_\mathrm{O}+f_\mathrm{O}}}{1+\frac{b_\mathrm{HC}}{b_\mathrm{OC}}f_\mathrm{O}}
\end{equation}

with $\mu_{i}=m_{i}/m_\mathrm{H}$, the binary diffusion parameters $b$, and the parameter
\begin{equation}\label{eq:phi}
	\Phi_{i}=\frac{L_\mathrm{H} kT}{3\pi GMm_{i} b_{\mathrm{H}i}}
\end{equation}
which represents approximately the ratio of drag to gravity (drag dominates if $\Phi_{i}>(\mu_{i-1})/\mu_{i}$). The factor $3\pi$ stems from
our adopted solid angle over which we assume that escape takes place and which is therefore included in the values of $L_\mathrm{H}$.
The binary diffusion parameter of O in H $b_\mathrm{HO}=4.8\times10^{17}T^{0.75}\,\mathrm{cm^{-1}\,s^{-1}}$ was taken from Table 1 of Zahnle and Kasting (1986).
$b_\mathrm{HC}$ was assumed to be equal to $b_\mathrm{HO}$, and $b_\mathrm{OC}$
is roughly estimated as $2\times10^{17}T^{0.75}\,\mathrm{cm^{-1}\,s^{-1}}$. However, we note that changing these parameters, as well as the adopted
temperature, does not affect the results if the hydrogen escape rate is large.

Eqs. \ref{eq:LO} and \ref{eq:LC} were derived under the assumption that the flow is isothermal and subsonic (Zahnle and Kasting, 1986), which is
actually not valid during the phase of saturated solar XUV emission studied here. However, they showed that these simpler analytic approximations
become comparable to the non-isothermal transonic solutions if $x_{i}\gg 1/\mu_{i}$ and $\Phi_i$ is large. These conditions are both fulfilled here
because the masses of O and C are much larger than H and hydrogen escapes very efficiently (hence, $\Phi_{i}\gg$). It was also assumed that the mixing
ratios $f_i$ are approximately constant with height. Expressions for $x_{i}$ without this constraint include terms with an exponential function that
goes to zero for large $\Phi_{i}$ (Zahnle and Kasting, 1986) and would therefore vanish for the cases studied here.

Figs. 7 to 10 show the temporal evolution of the partial surface pressures of H, O, and C normalized
to the initial total surface pressure for the four cases of the outgassed
atmospheres given in
Table 2. These results have been obtained by adopting the modeled hydrogen loss rates shown in the cases CI, CII, CIII and CIV in Table 3
corresponding to 100 times the present solar XUV flux and a lower boundary temperature $T_0$ of 200 K but
low and high heating efficiencies $\eta$ of 15 \% and 40\%, and $z_0$ at 100 km and 1000 km altitude.
This temperature is also used for evaluating $\Phi_{i}$, but choosing a different value
does not affect the results because the large $L_\mathrm{H}$ dominates. The initial hydrogen inventory evolves with a constant escape rate, because the timescale for
total hydrogen loss occurs during a time frame between $\sim$ 0.4 to 12 Myr, well below the time it takes the Sun to drop out of its saturation phase.
The evolution of the inventories, and hence
partial surface pressures, of O and C are found numerically by integration of eqs. \ref{eq:LO} and \ref{eq:LC}.

From these figures one can see that the timescale for
complete loss of H, O, and C, for a $\sim 50$ bar H$_2$O and $\sim 10$ bar CO$_2$ atmosphere for low $\eta$ and $z_0$ occurs
in less than 2.5 Myr. If the base of the thermosphere would expand from 100 km to 1000 km
and $\eta=40$ \%, such a steam atmosphere would be lost after $\sim$ 0.4 Myr. Depending on the initial volatile content
and assumed heating efficiencies and $z_0$, steam atmospheres with $\sim$ 260 bar H$_2$O and $\sim$ 55 bar CO$_2$
would be lost from early Mars between $\sim$2.1 and 12 Myr.
We also note that a magnetosphere would not protect the escape of the bulk atmosphere under these conditions because most of the atoms
escape as neutrals until they become ionized due to the interaction with the
early solar wind and plasma environment at large planetary distances (Kislyakova et al., 2013; Lammer, 2013).

As discussed above, the time scale for cooling of the steam atmosphere to temperature-pressure values, that water can condense and
build lakes or even oceans is very important and the influence of energy deposition on planetary surfaces by frequent impacts of large planetsimals or
small embryos has to be studied in coupled magma ocean-protoatmosphere models in the future. Therefore, it is also possible,
that all of our studied steam atmosphere scenarios presented in table 2 may have been lost within a few Myr, before the atmospheres cooled to
temperatures that big lakes or oceans could have formed.

However, for outgassed steam atmospheres with surface pressures $\gg 50$ bar, the timescale for total escape compared to the steam atmosphere cooling timescale
could be larger, so that large lakes or water oceans could have been formed sporadically. In such scenarios water condensed and
could have been present on the planet's surface for short time until the high thermal escape rates and impactors evaporated it again
(Genda and Abe, 2005). During this time and also during later stages a fraction of condensed, or via later impacts delivered H$_2$O,
may have been again incorporated by hydrothermal alteration processes such as serpentinization, so that remaining parts of it could
be stored even today in subsurface serpentine (Chassefi\`{e}re et al., 2013).

We point out that a detailed photochemical study, which includes
processes such as dissociation, ionization, etc. of the outgassed CO$_2$ molecules
is beyond the scope of the present study. Our expectation that no dense CO$_2$ atmosphere has build up on early Mars during the first 100 Myr is also
supported by a study of Tian et al. (2009), who showed that the thermal escape of
C atoms was so efficient even during the early Noachian, $>$4.1 Gyr ago, that a CO$_2$-dominated martian
atmosphere could not have been maintained, and Mars most likely has begun its origin colder. In agreement with Lammer et al. (2013a)
by the mid to late Noachian, as one can see from Fig. 4, the solar XUV flux would have become much weaker
allowing the build up of a secondary CO$_2$ atmosphere by volcanic outgassing (Grott et al., 2011) and/or impact delivered volatiles.

Our results are also in agreement with the conclusions of Bibring et al. (2005), which are based on the so far not detected carbonates,
that no major surface sink of CO$_2$ is present and the initial CO$_2$, if it was more abundant, should have been lost
from Mars very early other than being stored in surface reservoirs after having been dissolved in liquid water at the surface.
However, it should be noted that the accumulation of
a secondary outgassed CO$_2$ atmosphere and volatiles, which could have been delivered by later impacts is highly dependent
on less efficient atmospheric escape processes after the strong early hydrodynamic loss during the XUV-saturation
phase of the young Sun as well as by the efficiency of carbonate precipitation, and
serpentinization during the Hesperian and Amazonian epochs (e.g., Chassefi\`{e}re and Leblanc, 2011a; 2011b; Lammer et al., 2013a; Niles et al., 2013) .

Our result that Mars lost most likely the majority of its initial H$_2$O inventory very early is in
support of the hypothesis presented by Albar\`{e}de and Blichert-Toft (2004) that the planet could not develop an efficient plate tectonic regime
due to
the rapid removal of water by hydrodynamic escape. These authors suggest that
the resulting low abundance of the remaining water in the martian mantle
combined with weaker gravity than on Earth acted
against the bending and foundering of lithospheric
plates and the planet instead took the dynamic route of
developing a thick stagnant lid. Because of the low size and gravity of Mars
not enough water could be incorporated into the Martian mantle before it was lost to space so that plate tectonics never began.

\section{Conclusions}
The production and loss of the earliest martian atmosphere
which consisted of captured nebula gas (H, He, etc.) and outgassed and impact delivered
volatiles (e.g. H$_2$O, CO$_2$, CH$_4$, etc.) have been studied. By using the latest knowledge of the origin
of Mars summarized in Brasser (2013), we estimated the protoatmosphere masses and partial pressures
and applied a 1-D hydrodynamic upper atmosphere model to the extreme XUV conditions of the
young Sun. Depending on the amount of the outgassed volatiles, as well as
the assumed heating efficiency and altitude location of the lower thermosphere, our results indicate
that early Mars lost its nebular captured hydrogen envelope and catastrophically outgassed steam atmosphere
most likely within $\sim$0.4 - 12 Myr by hydrodynamic escape of atomic
hydrogen. The main reasons for the fast escape
of even a steam atmosphere with an amount of $\sim$70\% of an Earth ocean and $\sim 50$ bar CO$_2$ within
$<$ 12 Myr are Mars' low gravity and the $\sim$100 times higher XUV flux of the young Sun, which lasted $\sim$100
Myr after the Solar Systems origin. The efficient escape of atomic hydrogen, drags heavier atoms within the escaping bulk atmosphere so that
they can also be lost to space. Our results support the hypotheses of Tian et al. (2009)
that early Mars could not build up a dense CO$_2$ atmosphere during the early Noachian. The results are
also in agreement with the hypothesis presented in Lammer et al. (2013a) that after the planet lost its protoatmosphere
the atmospheric escape rates were most likely balanced with the volatiles, which have been outgassed by volcanic activity
and delivered by impacts until the activity of the young Sun decreased,
so that the atmospheric sources could dominate over the losses $\sim 4.2-3.8$ Gyr ago.

\textbf{Acknowledgments}\\
P. Odert, H. Lammer, K. G. Kislyakova and Yu. N. Kulikov acknowledge support from the Helmholtz Alliance project
``Planetary Evolution and Life''. E. Dorfi, M. G\"{u}del, K. G. Kislyakova, H. Lammer, A. St\"{o}kl and E. A. Dorfi
acknowledge the Austrian Science Fund (FWF) for supporting this study via the FWF NFN project S116 ``Pathways to Habitability:
From Disks to Active Stars, Planets and Life'', and the related FWF NFN subprojects, S 116 02-N1
``Hydrodynamics in Young Star-Disk Systems'', S116 604-N16 ``Radiation \& Wind Evolution from T Tauri Phase to ZAMS and Beyond'',
and S11607-N16 ``Particle/Radiative Interactions with Upper Atmospheres
of Planetary Bodies Under Extreme Stellar Conditions''. M. Leitzinger and P. Odert acknowledge also support from the FWF project P22950-N16.
N. V. Erkaev acknowledges support by the RFBR grant No 12-05-00152-a. Finally, H. Lammer thanks M. Ikoma from the
Department of Earth and Planetary Science, of the University of Tokyo, Japan, for discussions related to
the accumulation of nebular-based hydrogen envelopes around Mars-mass bodies. Finally the authors thank an anonymous referee
for the interesting and important suggestions and recommendations that helped to improve the results of our study.

\end{document}